\documentclass[11pt]{article}
\usepackage[letterpaper, portrait, margin=1in]{geometry}

\usepackage{mathtools}

\usepackage{bm}

\usepackage{enumerate}

\usepackage{color}

\def\blue{\color{blue}}

\def\XXint#1#2#3{{\setbox0=\hbox{$#1{#2#3}{\int}$}
     \vcenter{\hbox{$#2#3$}}\kern-.5\wd0}}

\def\Tr{\mbox{Tr}\,}
\def\Det{\mbox{Det}\,}
\def\tr{\mbox{tr}\,}
\def\sgn{\,\mbox{sgn}\,}
\newcommand{\R} {\mbox{Re}\,}

\newcommand{\la}{\label}
\newcommand{\be}{\begin{equation}}
\newcommand{\ee}{\end{equation}}
\newcommand{\bea}{\begin{eqnarray}}
\newcommand{\eea}{\end{eqnarray}}

\newcommand{\p}{\partial}
\newcommand{\ba}{\begin{align}}
\newcommand{\ea}{\end{align}}

\begin{document}


\title{\bf Topology, geometry and quantum interference in condensed matter physics}

\author{Alexander G. Abanov \medskip \\
Department of Physics and Astronomy and \\ Simons Center for Geometry and Physics, 
\\Stony Brook University, \\ Stony Brook, NY 11794, USA }

\maketitle

\pagenumbering{roman}

\begin{abstract}
The methods of quantum field theory are widely used in condensed matter physics. In particular, the concept of an effective action was proven useful when studying low temperature and long distance behavior of condensed matter systems. Often the degrees of freedom which appear due to spontaneous symmetry breaking or an emergent gauge symmetry, have non-trivial topology. In those cases, the terms in the effective action describing low energy degrees of freedom can be metric independent (topological). We consider a few examples of topological terms of different types and discuss some of their consequences. We will also discuss the origin of these terms and calculate effective actions for several fermionic models. In this approach, topological terms appear as phases of fermionic determinants and represent quantum anomalies of fermionic models. In addition to the wide use of topological terms in high energy physics, they appeared to be useful in studies of charge and spin density waves, Quantum Hall Effect, spin chains, frustrated magnets, topological insulators and superconductors, and some models of high-temperature superconductivity. \\

\medskip
\noindent
These notes are based on the lectures given by the author at ``SERC School on Topology and Condensed Matter Physics'' in Kolkata, India in December 2015. 
\end{abstract}

\tableofcontents			

\pagenumbering{arabic}

\newcounter{prn}
\newcommand{\prn}[1]{\refstepcounter{prn}\label{#1}}


\section{Introductory remarks}

\subsection{Theory of Everything in condensed matter physics}

In condensed matter physics, we believe that we know the ``Theory of Everything'' -- the fundamental equations potentially describing all observable phenomena in condensed matter physics. Essentially, those equations are Schr\"odinger equations for electrons and nuclei, together with Maxwell equations describing electromagnetic interactions. \cite{laughlin2000theory} However, there is a long way from knowing fundamental equations and being able to actually describe collective behavior of $10^{20}$ or so nuclei and electrons forming condensed matter systems, like, liquids, solids, superfluids, superconductors, quantum Hall systems etc. Having ``more'' particles makes macroscopic systems behave very differently from collections of just a few particles. New qualitative features appear when one goes from microscopic to macroscopic systems. \cite{anderson1972more} We refer to new phenomena appearing at macroscopic scales as to ``emergent phenomena''. 

The goal of condensed matter physics is not finding fundamental laws but rather finding their consequences. In particular, we are interested in finding efficient ways to describe emergent macroscopic phenomena. While it is very hard to derive macroscopic phenomena by solving fundamental equations we have a few guiding principles that allow us to write effective descriptions of those phenomena. Such principles include the use of symmetries and associated conservation laws, mechanisms of spontaneous symmetry breaking, the concept of quasiparticles etc. Early examples of effective descriptions include thermodynamics and hydrodynamics. 

In these lectures, we focus on the ``topological properties'' of condensed matter systems and their descriptions. Topological properties, in general, are the properties robust with respect to continuous deformations. They are emergent properties and it is important to understand them in the context of condensed matter physics as they might be the most stable properties insensitive to deformations and perturbations always present in realistic materials. The main focus of these lectures will be on topological properties related to quantum physics.

\subsection{Spontaneous symmetry breaking and an emergent topology}

If there were no separation of scales in Nature, the task of 
theoretical physicists would be formidable. Fortunately, in many cases one can ``integrate out'' fast degrees of freedom and effectively describe properties of microscopic systems at low 
temperatures, low frequencies, and large distances using relatively 
simple continuous field theory descriptions. This happens due to 
the presence of exact or approximate symmetries in the underlying 
microscopic system. More precisely, it is due to a phenomenon of spontaneous 
symmetry breaking.

Suppose that the exact Hamiltonian of some condensed matter system has
some continuous symmetry given by Lie group $G$.  A good system to keep
in mind as an example is an isotropic ferro- or antiferromagnet with
an SU(2) symmetry with respect to global rotations of all spins.  Then
it is possible that at some values of parameters of the
Hamiltonian,\footnote{We consider here the case of zero temperature for
simplicity.} the ground state of the system breaks the symmetry up to
some subgroup $H$ of $G$.  If this happens, we say that the symmetry of
the Hamiltonian is spontaneously broken by its ground state.  One can
characterize this ground state by some element $n$ of a coset space
$G/H$.  In our example of the magnet we take $H=SO(2)=U(1)$ and
$G/H=SU(2)/U(1)=S^{2}$.  The element of a coset space in this case is
a point of two-dimensional sphere $S^{2}$ which labels the direction
of magnetization of our system and subgroup $H$ is just the group of
all rotations around the direction of magnetization which is obviously
a symmetry of the Hamiltonian and of the ground state of the system. 
In the presence of spontaneous breaking of a continuous symmetry the
ground state is infinitely degenerate, since any $n\in G/H$ gives the
ground state with the same energy.  Indeed, any two states
characterized by $n_{1},n_{2}\in G/H$ have the same energy since they can
be transformed into one another by some element $g\in G$, which is an
exact symmetry transformation of the Hamiltonian.

Now consider another state of the quantum system which 
locally, in the vicinity of spatial point $x$, is very close to the ground state of 
the system labeled by some $n(x)\in G/H$. We assume that $n(x)$ is not 
constant in space but changes very slowly with a typical wavenumber $k$. 
In such a case, $n(x)$ is called an order parameter of the system.
We denote the energy of this state per unit volume measured from the 
energy of the ground state $\epsilon(k)$. The 
limit of small $k\to 0$ corresponds to the order parameter which is 
constant in space $n(x)=n_{0}$ and, therefore, $\epsilon(k)\to 0$ as 
$k\to 0$. We obtain that when continuous symmetry is spontaneously 
broken, the ground state of the system is not isolated but there are 
always excited states whose energies are infinitesimally close to the 
ground state energy. These heuristic arguments can be made more 
rigorous and lead to the Goldstone theorem\footnote{For a full 
formulation of the Goldstone theorem for relativistic field theory as well 
as for its proof see e.g., 
\cite{PeskinSchroeder}. We avoid it here because we are generally interested 
in a wider range of 
systems, e.g., without Lorentz invariance. There are still some analogs 
of the Goldstone theorem there. 
For example in the case of a ferromagnet there are still 
massless particles -- magnons. However, the number of independent 
massless particles is not correctly given 
by the Goldstone theorem for relativistic systems.}. The theorem states that in 
quantum field theory with spontaneously broken continuous symmetry 
there are massless particles which energy $\epsilon(k)\to 0$ as $k\to 
0$. 

If one is interested in low energy physics one necessarily should take 
these massless modes (or Goldstone bosons) into account. Moreover, the 
nature of these massless modes is dictated essentially by the 
symmetry (and its breaking) of the system, and one expects, therefore, 
that the correct low energy description should depend only on 
symmetries of the system but not on its every microscopic detail. 

A natural variable describing the dynamics of Goldstone modes is 
the order parameter itself. For example, in the case of relativistically invariant 
system described by the order parameter $\mathbf{n}\in S^{2}$, we 
immediately write
\be
    S_{\rm NLSM} =\int d^{d+1}x\, 
    \frac{1}{2g}(\partial_{\mu}\mathbf{n})^{2} + (\mbox{other terms}).
 \label{nlsm}
\ee
We have written here the most obvious term of an effective action
which is both Lorentz invariant and SU(2) invariant (with respect to
rotations of a unit, three-component vector $\mathbf{n}^{2}=1$).  Here
$g$ is a coupling constant which should be obtained from a detailed
microscopic theory.  The ``other terms'' are the terms which are
higher order in gradients and (possibly) topological terms.  The model
(\ref{nlsm}) is referred to as a ``non-linear
$\sigma$-model''\footnote{The origin of the term is in effective
theories of weak interactions \cite{gell1960axial,1968-AdlerDashen}.  Non-linear comes from the
non-linear realization of symmetries in this model.  E.g., constraint
$\mathbf{n}^{2}=1$ is non-linear.  Sigma ($\sigma$) is a historic
notation for the ``order parameter'' in theories of weak
interactions.}.  In different spatial dimensions, higher gradient terms
of non-linear $\sigma$-models might be relevant.  We are not
discussing those terms as well as the issue of renormalizability 
of $\sigma$-models, concentrating instead on the allowed topological
terms.  Therefore, we will keep only the kinetic term
$\frac{1}{2g}(\partial_{\mu}\mathbf{n})^{2}$ in the gradient expansion of
an effective Lagrangian as well as all allowed topological terms.

Before proceeding to our main subject -- topological terms, let us 
make two important remarks. 
Firstly, very often (especially in condensed matter systems) the symmetries of the Hamiltonian are 
approximate and there are terms in the Hamiltonian which explicitly 
but weakly break the symmetry. This does not invalidate the speculations of this section. 
The difference will be that would-be-Goldstone particles acquire small 
mass. The weaker the explicit symmetry breaking of the Hamiltonian the 
smaller is the mass of ``Goldstone'' particles. One can proceed with the 
derivation of the non-linear $\sigma$-model which will contain weak 
symmetry breaking terms (such as easy-axis anisotropy for magnets). 
This model will have non-trivial dynamics at energies bigger than the 
smallest of masses.

Second remark is that there are other mechanisms in addition to the
spontaneous symmetry breaking which result in low energy excitations. 
One of the most important mechanisms is realized when local (or gauge) 
symmetry is present. Then, gauge invariance plus locality demands the 
presence of massless particles (e.g, photons) in the system. 
The low energy theories in this case are gauge theories. Similar 
to an explicit symmetry breaking in case of Goldstone particles there 
are mechanisms which generate masses for gauge bosons. These are, 
e.g., Higgs mechanism and confinement of gauge fields. We will have 
some examples of topological terms made out of gauge fields in these 
lectures although our main focus will be on non-linear sigma 
models\footnote{Non-linear sigma models and gauge theories have a lot 
in common \cite{PolyakovBook-1987}.}. We also do not consider here cases with massless fermionic 
degrees of freedom we concentrate exclusively on bosonic effective 
theories. 

\subsection{Additional reading}

The focus of these lectures is on the effect of topological terms in the action on physical properties of condensed matter systems. We are not discussing here classification of topological defects in textures in ordered media. The latter is a well developed subject (see the classical review \cite{1979-Mermin}). For reader's convenience we collected a few exercises on topological textures and related examples in Appendix~\ref{app:topdefects}.

The subject of topological terms or broader ``topological phases of matter'' is huge, and we do not do it justice in these lectures. In particular, I do not try to give a complete bibliography in these lectures. Instead, with a few exceptions I refer not to the original papers but to textbooks or reviews.

Homotopy classification of topological defects and textures in ordered media is not discussed in this lectures. However, it is a necessary prerequisite to understanding topological terms discussed here. I recommend a classic reference \cite{1979-Mermin}.  To make this text more self-contained I also collected few exercises on that topic in Appendix~\ref{app:topdefects} and relevant homotopy groups in Appendix~\ref{app:homgroups}. 

I would recommend the following textbooks close in spirit to the point of view presented here \cite{PolyakovBook-1987,2010-AltlandSimons,fradkin2013field}. Some of the technical details of fermionic determinant calculations can be found in \cite{AbanovWiegmann-2000,abanov2000hopf}. 
Topological terms are intimately related to geometric or Berry phases \cite{1989-ShapereWilczek} and to quantum anomalies in field theories \cite{1985-TJZW}. 

In these lectures, I avoid using any advanced topological and geometrical tools. However, I highly recommend studying all necessary mathematics seriously. There are many beautiful books that give good introduction to the subject for physicists. See, for example Refs.~\cite{1985-DFN,nakahara2003geometry,monastyrsky2013topology,1991-Arnold,stone2009mathematics}.

\section{Motivating example: a particle on a ring.}

\subsection{Classical particle on a ring: Action,  Lagrangian, and Hamiltonian}

As a simple motivating example let us consider a particle on a ring. Classically, the motion can be described by the principle of least action. A classical action $S$ of a particle can be taken as
\bea
	S[\phi] &=& \int dt\, L(\phi,\dot\phi),
 \\
 	L &=& \frac{M}{2}\dot\phi^{2} +A\dot\phi,
 \la{por-A}
\eea
should be minimal (locally) on classical trajectories. Here, the angle $\phi(t)$ is chosen to be a generalized coordinate of the particle on a ring, $M$ is a moment of inertia of a particle (or mass for a unit ring), $A$ is some constant. 

Euler-Langrange equations of motion are given in terms of Lagrangian $L$ by $\frac{d}{dt}\frac{\partial L}{\partial\dot\phi}-\frac{\partial L}{\partial \phi}=0$, or explicitly
\be
	M\ddot\phi=0.
 \la{por-eqm}
\ee
A particle, given an initial velocity, moves with constant angular velocity along the ring. Notice, that the last term of (\ref{por-A}) does not have any effect on the motion of the particle. Indeed, this term is a total time derivative and can not affect the principle of least action \cite{landau1976mechanics}. 

Given an initial position of a particle on a ring at $t=t_{1}$ and a final position at $t=t_{2}$ there are infinitely many solutions of (\ref{por-eqm}). They can be labeled by the integer number of times particle goes around the ring to reach its  final position. This happens because of \textit{nontrivial topology} of the ring -- one should identify $\phi=\phi+2\pi$ as labeling the same point on the ring. This 
 is not very important classically as we can safely think of the angle $\phi$ taking all real values from $-\infty$ to $+\infty$. Given initial position $\phi(t_{1})=\phi_{1}$ and initial velocity $\dot\phi(t_{1})=\omega_{1}$ one can unambiguously determine the position of the particle $\phi(t)$ at all future times using (\ref{por-eqm}).

Let us now introduce the momentum conjugated to $\phi$ as
\be
	p = \frac{\partial L}{\partial \dot\phi} = M\dot\phi +A,
\ee
and the Hamiltonian as
\be
	H = p\dot\phi -L = \frac{1}{2M}(p-A)^{2}.
 \la{por-cH}
\ee
Corresponding Hamilton equations of motion 
\bea
	\dot\phi &=& \frac{1}{M}(p-A),
 \\
 	\dot p &=& 0,
\eea
are equivalent to (\ref{por-eqm}).

Notice that although the parameter $A$ explicitly enters Hamiltonian formalism, it only changes the definition of generalized momentum $M\dot\phi +A$ instead of more conventional $M\dot\phi$. It does not change the solution of equations of motion and can be removed by a simple canonical transformation $p \to p+A$. We will see below that this changes for a \textit{quantum} particle.

\subsection{Quantum particle on a ring: Hamiltonian and spectrum}

Let us now consider a quantum particle on a ring. We  replace classical Poisson's bracket $\{p,\phi\}=1$ by quantum commutator $[p,\phi]=-i\hbar$ and use $\phi$-representation, i.e., we describe our states by wave functions on a ring $\psi(\phi)$. In the following, we will put $\hbar=1$. In this representation, we can use $p =-i\partial_{\phi}$ and rewrite (\ref{por-cH}) as a quantum Hamiltonian
\begin{equation}
     H = \frac{1}{2M}\left(-i\partial_{\phi}-A\right)^{2}.
  \label{eq:plroth}
\end{equation}
The eigenstates and eigenvalues of this Hamiltonian are given by solutions of stationary Schr\"odinger equation $H\psi = E\psi$. We impose \textit{periodic boundary conditions} requiring $\psi(\phi+2\pi)=\psi(\phi)$, i.e., the wave function is required to be a single-valued function on the ring.	
The eigenfunctions and eigenvalues of (\ref{eq:plroth}) are given by
\begin{eqnarray}
      \psi_{m} &=& e^{im\phi}, 
  \\
       E_{m} &=& \frac{1}{2M}(m-A)^{2},
 \label{eq:plrotsp}
\end{eqnarray}
\vspace{0.5cm}
where $m=0,\pm 1,\pm 2,\ldots$ is any integer number - the quantized eigenvalue of the momentum operator $p=-i\partial_{\phi}$. We notice that although the classical model is not sensitive to the parameter $A$, the quantum one is, because of the quantization of $p$. The parameter $A$ can be interpreted as a vector potential of the magnetic flux penetrating the ring. This vector potential is not observable in classical mechanics but affects the quantum spectrum because of multiple-connectedness of the ring (there are many non-equivalent ways to propagate from the point 1 to the point 2 on a ring). More precisely our parameter $A$ should be identified with the vector potential multiplied by $\frac{e}{\hbar c}$. It corresponds to the magnetic flux through the ring $\Phi=A \Phi_{0} $, where $\Phi_{0}$ is a flux quantum $\Phi_{0}=2\pi \frac{\hbar c}{e}$. 

The $A$-term of the classical action -- \textit{topological term} -- can be  written as
\be
	S_{top} = \int_{t_{1}}^{t_{2}}dt\, A\dot\phi 
	= 2\pi A \frac{\phi_{2}-\phi_{1}}{2\pi}= \theta \frac{\Delta\phi}{2\pi}.
 \la{ringtop}
\ee
It depends only on the initial and final values $\phi_{1,2}=\phi(t_{1,2})$ and changes by $\theta =2\pi A$ every time the particle goes a full circle around the ring in counterclockwise direction. The conventional notation $\theta$ for a coefficient in front of this term  gave  the name \textit{topological theta-term} for this type of topological terms.

The spectrum (\ref{eq:plrotsp}) is shown in Figure~\ref{fig:porspectrum} for three values of flux through the ring: $\theta = 0,\pi,\pi/2$ ($A=\Phi/\Phi_{0}=0,1,1/2$). 
\begin{figure}
\bigskip
\begin{center}
 \includegraphics[width=4cm]{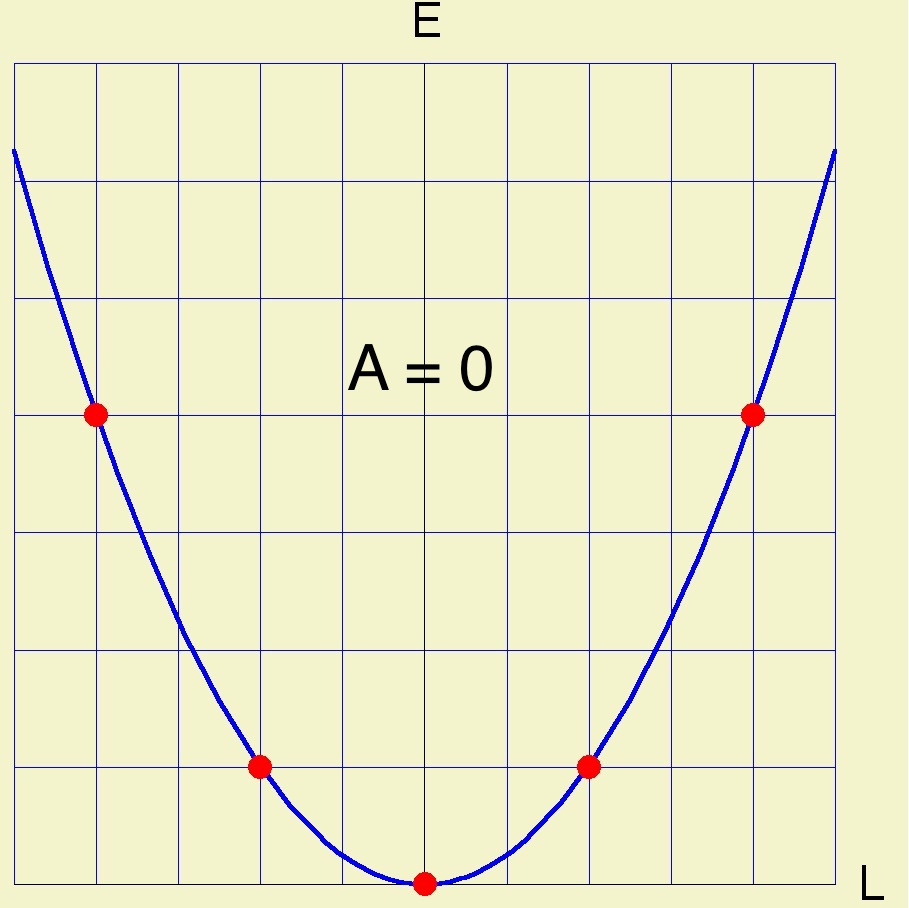} \quad
  \includegraphics[width=4cm]{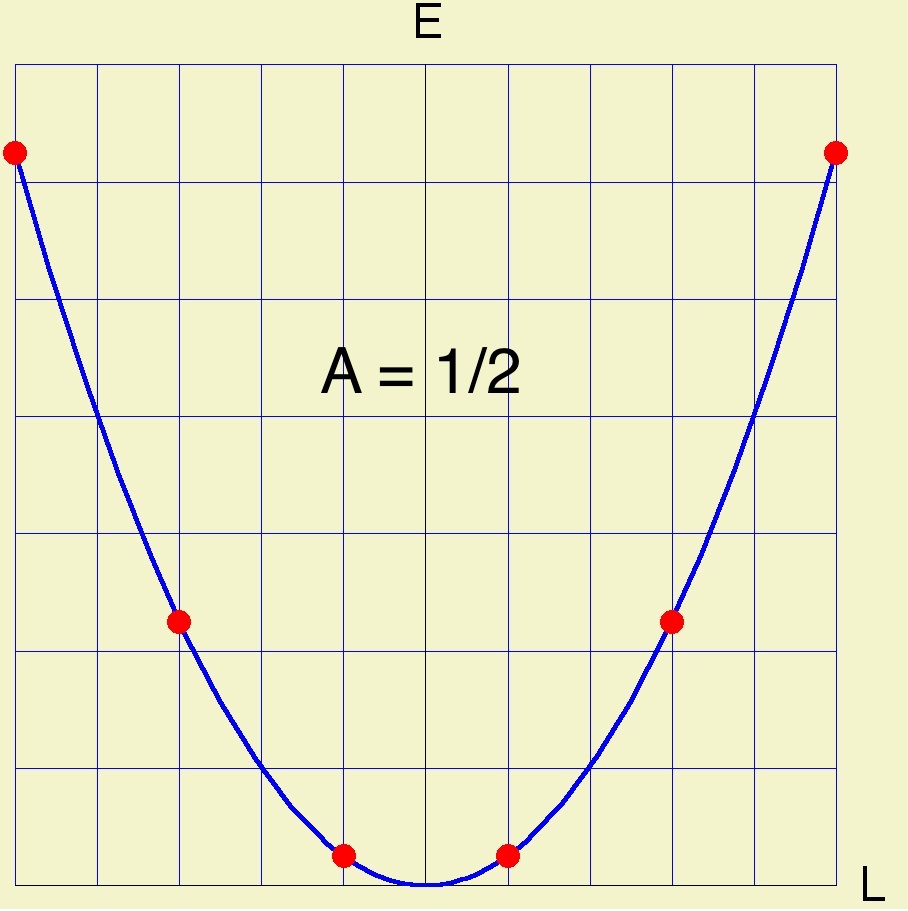} \quad
    \includegraphics[width=4cm]{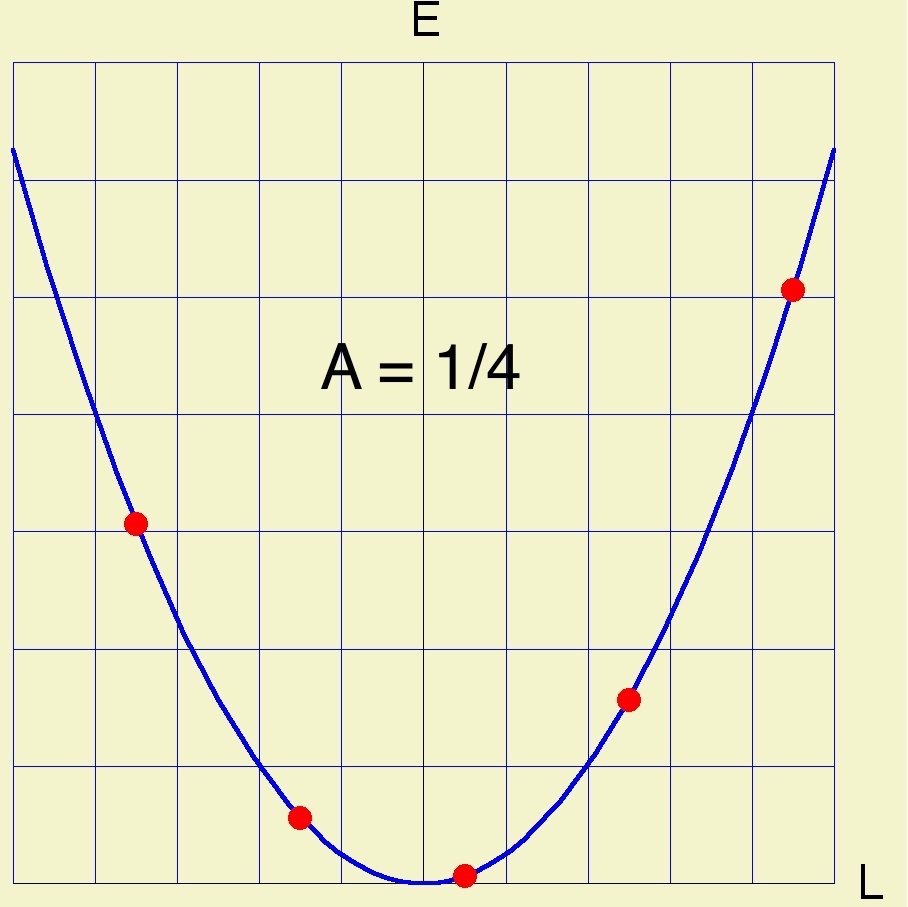}
\end{center}
\caption{The spectrum of the particle on a ring is shown for $A = \theta/2\pi=0,1/2,1/4$ respectively. The classical energy $E(p)$ is represented by a parabola and does not depend on the parameter $A$.}
\la{fig:porspectrum}
\end{figure}

Several comments are in order. (i) An integer flux $A$-integer or $\theta$ - multiple of $2\pi$ does not affect the spectrum. (ii) There is an additional symmetry (parity) of the spectrum when $\theta$ is a multiple of $\pi$ (integer or half-integer flux). (iii) For half-integer flux $\theta=\pi$, the ground state is doubly degenerate $E_{0}=E_{1}$.

Finally, let us try to remove the $A$ term by canonical transformation as in the classical case. We make a gauge transformation $\psi \to e^{iA\phi}\psi $ and obtain $p \to p+A$ and $H=\frac{1}{2M}(-i\partial_{\phi})^{2}$. One might think that we removed the effects of the $A$ term completely. However, this transformation changes the boundary conditions of the problem replacing them by \textit{twisted boundary conditions} $\psi(\phi+2\pi)= e^{-i 2\pi A}\psi(\phi)$. The eigenfunctions satisfying twisted boundary conditions are $\psi_{m}=e^{i(m-A)\phi}$ and produce the same eigenvalues (\ref{eq:plrotsp}). We conclude that it is not possible to remove the effects of topological $A$-term in quantum mechanics. The parameter $A$ can be formally removed from the Hamiltonian by absorbing it into the boundary conditions. This, however, does not change the spectrum and other physical properties of the system.

\subsection{Quantum particle on a ring: path integral and Wick's rotation}

Quantum mechanics of a particle on a ring described by the classical action (\ref{por-A}) can be represented by path integral
\be
	Z = \int D\phi\; e^{iS[\phi]},
 \la{por-Z}
\ee
where integration is taken over all possible trajectories $\phi(t)$ (with proper boundary values).
In this approach the contribution of the topological term  to the weight in the path integral is the phase
$e^{i\theta \Delta\phi/(2\pi)}$ which is picked up by a particle moving in the presence of the vector potential.

Let us perform \textit{Wick's rotation} replacing the time by an imaginary time $\tau=it$. Then
\bea
	\int dt\, \frac{M\dot\phi^{2}}{2} & \to & i \int d\tau\, \frac{M\dot\phi^{2}}{2},
 \\
 	\int dt\, A\dot\phi & \to & \int d\tau\, A\dot\phi,
 \la{por-topWick}
\eea
where in the r.h.s dot means the derivative with respect to $\tau$. The path integral (\ref{por-Z}) is then replaced by a Euclidean path integral 
\begin{equation}
     Z=\int_{e^{i\left[\phi(T)-\phi(0)\right]}=1} 
     {\cal D}\phi\, e^{-S\left[\phi\right]},
  \label{eq:plrotz}
\end{equation}
where the action
\begin{equation}
     S=\int_{0}^{\beta}d\tau\, \left[\frac{M}{2}\dot\phi^{2} -iA\dot\phi\right].
  \label{eq:plrots}
\end{equation}
We considered the amplitude of the return to the initial point in time $\beta$, i.e. $0<\tau<\beta$. This requires periodic boundary conditions in time 
$e^{i\phi(0)}=e^{i\phi(\beta)}$. 

We notice here that because the $A$-term is linear in time derivative it does not change its form under Wick's rotation (\ref{por-topWick}) and therefore,  is still imaginary in Euclidean formulation (\ref{eq:plrots}). Without imaginary term one could think about $e^{-S}$ as of the Boltzmann weight in the classical partition function.

One can satisfy the boundary 
conditions as $\phi(\beta)-\phi(0)=2\pi Q$ with any integer $Q$. We can rewrite 
the partition function (\ref{eq:plrotz})  as:
\begin{equation}
     Z=\sum_{Q=-\infty}^{+\infty}e^{i \theta Q} \int_{\phi(\beta)-\phi(0)=2\pi Q} 
     {\cal D}\phi\, e^{-\int_{0}^{\beta} d\tau\, \frac{M}{2}\dot\phi^{2} }.
  \label{eq:plrotzs}
\end{equation}
We notice here that $\theta =2\pi n$ -- multiple of $2\pi$ --  is equivalent to $\theta=0$. 
Second, we notice that the partition function is split into the sum of path 
integrals over distinct {\em topological sectors} characterized by 
an integer number $Q$ which is called the {\em winding number}. The 
contributions of topological sectors to the total partition function 
are weighed with the complex weights $e^{i \theta Q}$.

For future comparisons, let us write (\ref{eq:plrots}) in terms of 
a unit two-component vector
$\mathbf\Delta=(\Delta_{1},\Delta_{2})=(\cos\phi,\sin\phi)$, 
$\mathbf\Delta^{2}=1$.
\begin{equation}
     S=\int_{0}^{\beta}d\tau\, \left[\frac{M}{2}\dot{\mathbf\Delta}^{2} 
     -iA(\Delta_{1}\dot\Delta_{2}-\Delta_{2}\dot\Delta_{1})\right].
  \label{eq:plrotsD}
\end{equation}
This is the simplest ($0+1$)-dimensional $O(2)$ non-linear $\sigma$-model.

\subsection{Quantum doublet}
 \la{sec:QD}

Let us consider a particular limit of a very light particle on a circle $M\to 0$ in the presence of 
half of the flux quantum $A=1/2$, $\theta=\pi$.
With this flux, the ground state of the system is doubly degenerate $E_{0}=E_{1}$ and the rest of the spectrum is separated by the energies $\sim 1/M \to \infty$ from the ground state (\ref{eq:plrotsp}). At large $\beta$ (low temperatures) we can neglect contributions of all states except for the ground state. 

We write the general form of the ground 
state wave function as $\alpha |+1/2\rangle + \beta |-1/2\rangle $, where $|+1/2\rangle=\psi_{0}$ and $|-1/2\rangle=\psi_{1}$. 
The ground state space $(\alpha,\beta)$ coincides with 
the one for a spin $1/2$. One might say that 
(\ref{eq:plrotz}-\ref{eq:plrots}) with $M\to 0$ realize a path integral 
representation for the quantum spin $1/2$. This representation does not 
have an explicit $SU(2)$ symmetry. We will consider an $SU(2)$-symmetric 
path integral representation for quantum spins later.

Meanwhile, let us discuss some topological aspects of a 
plane rotator problem.

\subsection{Full derivative term and topology}

From a mathematical point of view, the motion of a particle on a 
unit circle  with periodic boundary conditions in time  is described 
by a mapping $\phi(\tau):\;\; S^{1}_{\tau}\to S^{1}_{\phi}$ 
of a circle formed by compactified time 
$S^{1}_{\tau}=\tau\in \left[0,\beta\right]$ into a circle $S^{1}_{\phi}=\phi\in 
\left[0,2\pi\right]$.
This mapping can be characterized by integer {\em winding number} $Q$ 
which tells us how many times the image $\phi$ goes around {\em 
target space} $S^{1}_{\phi}$ when variable $\tau$ changes from $0$ to 
$\beta$. 

It can be shown that two such mappings $\phi_{1}(\tau)$ and $\phi_{2}(\tau)$ can 
be continuously deformed into one another if and only if they have the same winding number. Therefore, all mappings are 
divided into topological classes enumerated by $Q=0,\pm 1,\pm 
2,\ldots$. Moreover, one can define a group structure on topological 
classes. First, we define the {\em product} of two mappings $\phi_{1}$ 
and $\phi_{2}$ as 
$$\phi_{2}\cdot\phi_{1}(\tau)=\left\{ \begin{array}{l}
       \phi_{1}(2\tau), \hspace{2.65cm} 
        \mbox{for }\;\;0<\tau<\beta/2 \,,
 \\
       \phi_{1}(\beta)+\phi_{2}(2\tau-\beta), \qquad \mbox{for }\;\;\beta/2<\tau<\beta \,.
       \end{array} \right. $$
If $\phi_{1}$ belongs to the topological class $Q_{1}$ and $ \phi_{2}$ to $Q_{2}$, 
their product belongs to the class $Q_{1}+Q_{2}$. One can say that the 
product operation on mappings induces the structure of Abelian group 
on the set of topological classes. In this case this group is the 
group of integer numbers with respect to addition. One can write this 
fact down symbolically as $\pi_{1}(S^{1})=Z$, where subscript one 
denotes that our time is $S^{1}$ and $S^{1}$ in the argument is our 
target space. One says that the {\em first (or fundamental) 
homotopy group of $S^{1}$ is the group of integers}.

There is a simple formula giving the topological class $Q\in Z$ in terms of $\phi(\tau)$
\be
	Q = \int_{0}^{\beta}\frac{d\tau}{2\pi}\, \dot\phi\,.
\ee

Let us now assume that we split our partition function into  the 
sum over different topological classes. What are the general 
restrictions on the possible complex weights which one can introduce 
in the physical problem. One can deform smoothly any mapping in the 
class $Q_{1}+Q_{2}$ into two mappings of classes $Q_{1}$ and $Q_{2}$ which are 
separated by a long time. Because of the multiplicative property of 
amplitudes, this means that the weights $W_{Q}$ associated with 
topological classes must form a (unitary) representation of 
the fundamental group of a target space. The 
only unitary representation of $Z$ is given by $W_{Q}=e^{i \theta Q}$ 
with $0<\theta<2\pi$ labelling different representations. In the case of plane 
rotator, these weights correspond to a phase due to the magnetic flux piercing the 
one-dimensional ring.

In more general case of, say, particle moving on the {\em manifold} 
$G$ (instead of $S^{1}$) we have to consider the fundamental group of 
the target space $\pi_{1}(G)$, find its unitary representations, and 
obtain complex weights which could be associated with different 
topological classes.

\subsection{Topological terms and quantum interference}

As it can be seen from (\ref{eq:plrotzs}) the presence of a topological term in the action ($\theta\neq 0$) results in the interference between topological sectors in the partition function. The Boltzmann weight calculated for a trajectory within a given topological sector $Q$ is additionally weighted with complex phase $e^{i\theta Q}$. This interference can not be removed by Wick's rotation.

\subsection{General definition of topological terms}

We \underline{define} generally \textit{topological terms} as \textit{metric-independent} terms in the action.

A universal object present in any local field theory is the symmetric \textit{stress-energy tensor} $T_{\mu\nu}$. It can be defined as a variation of the action with respect to the metric $g^{\mu\nu}$. More precisely, an infinitesimal variation of the action can be written as
\be
	\delta S = \int dx\,\sqrt{g}\; T_{\mu\nu}\delta g^{\mu\nu},
\ee
where $\sqrt{g}\,dx$ is an invariant volume of space-time.

It immediately follows from our definition of topological terms that they do not contribute to the stress-energy tensor. If in a field theory all terms are topological we have $T_{\mu\nu}=0$ for such a theory. These theories are called topological field theories.

A particular general covariant transformation is the rescaling of time. Topological terms do not depend on a time scale. Therefore, the corresponding Lagrangians are linear in time derivatives. They do not transform under Wick's rotation and are always imaginary in Euclidean formulation. They describe quantum interference which is not removable by Wick rotation.

\subsection{Theta terms and their effects on the quantum problem}

Theta terms are topological terms of a particular type. They appear when  there exist nontrivial topological textures in space-time. Essentially, these terms are just complex weights of different topological sectors in the path integration. We will go over more details on $\theta$-terms later in the course. 

In addition to being imaginary in Euclidean formulation as all other topological terms, $\theta$-terms have also some special properties. These properties  distinguish them from other types of topological terms. The following is a partial list of the features of topological $\theta$-terms and of their manifestations.
\begin{itemize}
	\item $\theta$-terms assign complex weights in path integral to space-time textures with integer topological charge $Q$
	\item Realize irreducible 1d-representations of $\pi_{D}(G)$, where $D$ is the dimension of space-time and $G$ is a target space
	\item Quantum interference between topological sectors
	\item Do not affect equations of motion
	\item Affect the spectrum  of a quantum problem by changing quantization rules
	\item Periodicity in coupling constant $\theta$ \footnote{We assume that configurations are smooth and the space-time manifold is closed (no boundary)}.
	\item $\theta$ is not quantized  (for $Q\in Z$)
	\item For $\theta=0,\pi$, there is an additional (parity) symmetry
	\item $\theta =\pi$ -- degeneracy of the spectrum. Gapless excitations.
	\item Equivalent to changes in boundary conditions.
	\item $\theta$ is a new parameter which appears from the ambiguity of quantization of the classical problem for multiply-connected configurational space. 
\end{itemize}

\subsection{Exercises}
\setcounter{prn}{0}

{\small

\prn{ex:por-path}
\subsubsection*{Exercise \ref{ex:por-path}: Particle on a ring, path integral}

The Euclidean path integral for a particle on a ring with magnetic flux through the ring is given by
$$
	Z = \int {\cal D}\phi\; e^{-\int_{0}^{\beta} d\tau\, \left(\frac{m\dot\phi^{2}}{2} 
	-i\frac{\theta}{2\pi}\dot\phi\right)}.
$$
Using the decomposition
$$
	\phi(\tau) = \frac{2\pi}{\beta}Q\tau +\sum_{l\in \mathbf{Z}}\phi_{l}e^{i\frac{2\pi}{\beta}l\tau},
$$
rewrite the partition function as a sum over topological sectors labeled by winding number $Q\in \mathbf{Z}$ and calculate it explicitly. Find the energy spectrum from the obtained expression.
\\

\textit{Hint}: Use summation formula
$$
	\sum_{n=-\infty}^{+\infty} e^{-\frac{1}{2}An^{2}+iBn} = \sqrt{\frac{2\pi}{A}}
	\sum_{l=-\infty}^{+\infty} e^{-\frac{1}{2A}(B-2\pi l)^{2}}.
$$

\prn{ex:por-1/2}
\subsubsection*{Exercise \ref{ex:por-1/2}: Spin 1/2 from a particle on a ring}

Calculate the partition function of a particle on a ring described in the previous exercise. Find explicit expressions in the limit $m\to 0$, $\theta\to \pi$ but  $\theta-\pi\sim m/\beta$. One can interpret the obtained partition function as a partition function of a spin 1/2. What is the physical meaning of the ratio $(\theta-\pi)/m$ in the spin 1/2 interpretation of the result?

\textit{Hint}: see Sec.~\ref{sec:QD}.

\prn{ex:por-metric}
\subsubsection*{Exercise \ref{ex:por-metric}: Metric independence of the topological term}

The classical action of a particle on a ring  is given by
$$
	S = \int dt_{p}\, \left(\frac{m\dot\phi^{2}}{2} 
	-\frac{\theta}{2\pi}\dot\phi\right),
$$
where $t_{p}$ is some ``proper'' time. 
Reparametrizing time as $t_{p}=f(t)$ we have $dt_{p} = f' dt$ and $dt_{p}^{2} ={f'}^{2}dt^{2}$ and identify the metric as $g_{00}={f'}^{2}$ and $g^{00}={f'}^{-2}$. We also have 
$\sqrt{g_{00}}=f'$. Rewrite the action in terms of $\phi(t)$ instead of $\phi(t_{p})$. Check that it has a proper form if written in terms of the introduced metric. Using the general formula for variation of the action with respect to a metric ($g=\mbox{det}\, g_{\mu\nu}$)
$$
	\delta S = \frac{1}{2}\int dx\, \sqrt{g}\; T_{\mu\nu}\delta g^{\mu\nu},
$$
find the stress-energy tensor for a particle on a ring. Check that $T_{00}$ is, indeed, the energy of the particle.

%

}

\section{Path integral for a single spin.}

Wess and Zumino introduced an effective Lagrangian to summarize the anomalies in current algebras \cite{1971-WessZumino}. E. Witten considered global (topological) aspects of this effective action \cite{1983-Witten-global}. Simultaneously, S. P. Novikov studied multi-valued functionals \cite{1982-Novikov}. The corresponding topological terms are referred to as Wess-Zumino-Novikov-Witten terms or more often as just Wess-Zumino terms. In this section, we consider the simplest quantum mechanical (0+1 dimensional) version of such a term which is relevant for path integral formulation of a quantum mechanics of a single spin.

\subsection{Quantum spin}

Let us consider a simple example of how Wess-Zumino effective Lagrangian appears from the ``current algebra''. To simplify the story we take an example of quantum spin $S$. This is a quantum mechanical system with an $SU(2)$ spin algebra playing the role of ``current algebra'' of quantum field theory. We have standard spin commutation relations
\be
 \la{crsu2}
    [S^a,S^b]=i\epsilon^{abc}S^c,
\ee
where $a,b,c$ take values $x,y,z$. We require that 
\be
    \mathbf{S}^2 = S(S+1),
\ee
where $2S$ is an integer number defining the representation (the value of spin). Let us consider the simplest possible Hamiltonian of a quantum spin in a constant magnetic field
\be
 \la{smf}
    H = -\mathbf{h}\cdot\mathbf{S}
\ee
and derive an operator equation of motion
\be
 \la{eqmop}
  \partial_t \mathbf{S} = i[H,\mathbf{S}] = -i[\mathbf{h}\cdot\mathbf{S},\mathbf{S}]
   =\mathbf{S}\times\mathbf{h}.
\ee
In the classical limit $S\to \infty$ (or $\hbar\to 0$) it is convenient to write $\mathbf{S}\to S\mathbf{n}$ so that $\mathbf{n}$ is a classical unit vector $\mathbf{n}^2=1$ and equation of motion(\ref{eqmop}) becomes classical equation of motion
\be
 \la{eqmcl}
  \partial_t \mathbf{n} 
   =\mathbf{n}\times\mathbf{h}.
\ee
The natural question immediately occurs is what classical action corresponds to this equation of motion. It turns out that writing down this action is not completely trivial problem if one desires for the action to have explicitly $SU(2)$ invariant form. 
Let us first derive it using non-invariant parameterization in terms of spherical angles $\mathbf n=(\sin\theta\cos\phi, \sin\theta\sin\phi, \cos\theta)$. We assume that the angle $\theta$ is  measured from the direction of magnetic field $\mathbf{h}=(0,0,h)$. The Hamiltonian (\ref{smf}) becomes $H=-hS\cos\theta$ and equations of motion (\ref{eqmcl}) become $\dot\phi=-h$ and $\dot\theta=0$ -- precession around the direction of magnetic field. We obtain these equations as Hamilton's equations, identifying the momentum conjugated to $\phi$ coordinate as
\be
	p_{\phi}= -S(1-\cos\theta).
\ee
Then the classical action of a single spin in magnetic field can be written as 
\be
 \la{cas}
    S[\mathbf{n}] = -4\pi S W_0 +\int dt\, S \mathbf{h}\cdot\mathbf{n},
\ee
where $W_0$ is defined using a particular choice of coordinates as 
\be
    W_0 = \frac{1}{4\pi} \int dt\,  (1-\cos\theta)\partial_t\phi
    =\frac{1}{4\pi} \int d\phi\,  (1-\cos\theta)
    =\frac{\Omega}{4\pi},
 \la{seff1loc}
\ee
where $\Omega$ is a solid angle encompassed by the trajectory of $\mathbf 
n(t)$ during time evolution. The first term in the action (\ref{cas}) has a form of $\int dt\, p_{\phi}\dot\phi$ and the second is a negative time integral of the Hamiltonian. 

Although (\ref{seff1loc}) has a nice geometrical meaning it is written in some particular coordinate system on two-dimensional sphere. It would be nice to have an expression for $W_{0}$ which is coordinate independent and explicitly $SU(2)$ invariant (with respect to rotations of $\mathbf{n}$). Such a form, indeed exists 
\be
    W_{0}=\int_{0}^{1}d\rho\, \int_{0}^{\beta} dt\, \frac{1}{8\pi}
    \epsilon^{\mu\nu}\mathbf n\cdot\left[\partial_{\mu}\mathbf n
    \times\partial_{\nu}\mathbf  n\right].
 \la{wz0}
\ee
Here we assume periodic boundary conditions in time $\mathbf{n}(\beta)=\mathbf{n}(0)$, $\rho$ is an auxiliary coordinate $\rho \in\left[0,1\right]$. 
$\mathbf n$-field is extended to $\mathbf n(t,\rho)$ in such a way that 
$\mathbf n(t,0)=(0,0,1)$ and $\mathbf n(t,1)=\mathbf n(t)$. Indices $\mu,\nu$ 
take values $t,\rho$.

\begin{figure}
   \centerline{\includegraphics[scale=0.5]{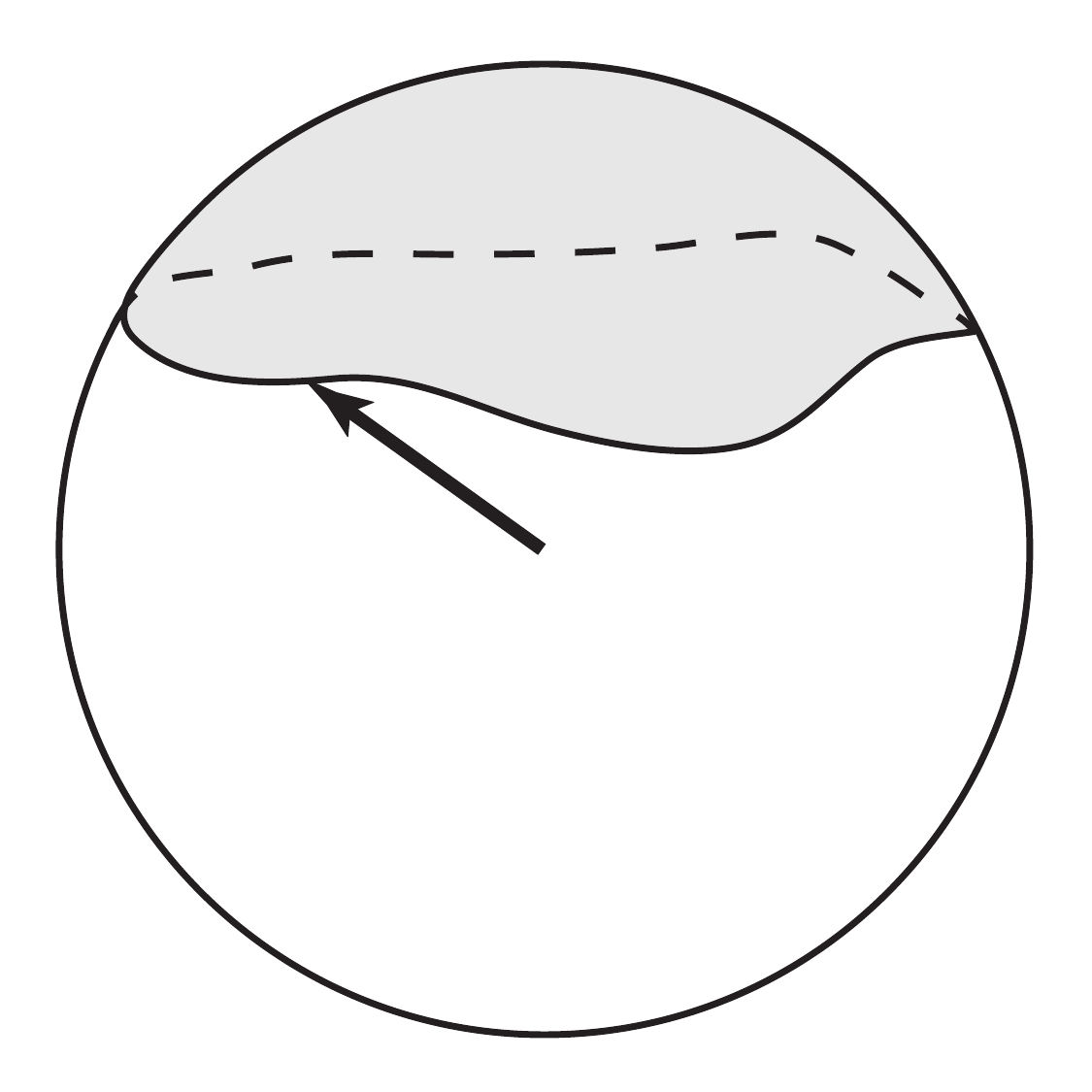}}
   \caption{\label{fig:WZ} The unit vector $\mathbf{n}(\tau)$ draws a 
   closed line on the surface of a sphere with unit radius during its 
   motion in imaginary time. Berry phase is proportional 
   to the solid angle
   (shaded region) swept by the vector $\mathbf{n}(t)$. One can 
   calculate this solid angle by extending $\mathbf{n}$ into the 
   two-dimensional domain ${\cal B}$ as $\mathbf{n}(\rho,t)$ and 
   calculating (\protect\ref{wz0}).}
\end{figure}

Wess-Zumino action (\ref{wz0}) has a very special property. 
Although it is defined as an integral over two-dimensional disk 
parameterized by $\rho$ and $t$ its variation depends only on the 
values of $\mathbf n$ on the boundary of the disk---physical time.
Indeed one can check that 
\bea
    \delta W_0 &=& \int_{0}^{1}d\rho\, \int_{0}^{\beta} dt\, \frac{1}{4\pi}
    \epsilon^{\mu\nu}\mathbf n\cdot\left[\partial_{\mu}\delta\mathbf n
    \times\partial_{\nu}\mathbf  n\right]
 \nonumber \\
    &=& \int_{0}^{1}d\rho\, \int_{0}^{\beta} dt\, \partial_{\mu} \left\{\frac{1}{4\pi}
    \epsilon^{\mu\nu}\mathbf n\cdot\left[\delta\mathbf n
    \times\partial_{\nu}\mathbf  n\right]\right\}
 \nonumber \\
    &=&  \frac{1}{4\pi}\int_0^\beta dt\, \delta\mathbf{n}
    \left[\dot{\mathbf{n}}\times\mathbf{n}\right],
 \la{WZvar}
\eea
where we used that $\delta\mathbf{n}\cdot\left[\partial_{\mu}\delta\mathbf n
    \times\partial_{\nu}\mathbf  n\right]=0$ because all three vectors $\delta\mathbf{n}$, $\partial_\mu \mathbf{n}$, and $\partial_\nu\mathbf{n}$ lie in the same plane (tangent to the two-dimensional sphere $\mathbf{n}^2=1$. Due to this property classical equation of motion does not depend on the arbitrary extension of $\mathbf{n}$ to $\rho \neq 1$.

\begin{figure}
   \centerline{\includegraphics[scale=0.5]{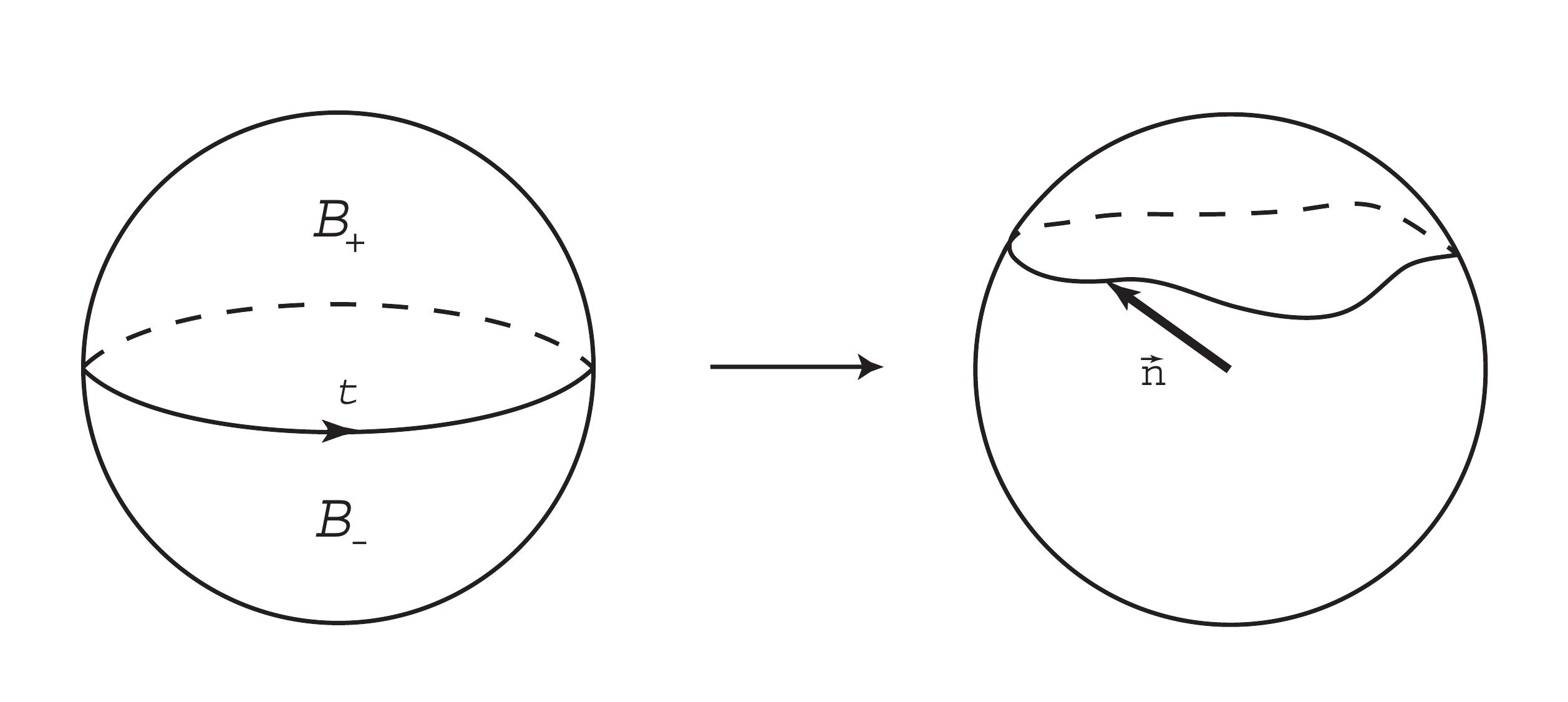}}
   \caption{\label{fig:WZmv} Two extensions $\mathbf{n}(t,\rho)$ and $\mathbf{n}'(t,\rho)$ 
   define a mapping $S^2\to S^2$. The difference $W_0[\mathbf{n}]-W_0[\mathbf{n}']$ gives
   a winding number of this mapping.}
\end{figure}

In quantum physics, however, not only the variation $\delta W_0$ but the weight $e^{2\pi iW_0}$ should not depend on unphysical  configuration $\mathbf{n}(t,\rho)$ but only on $\mathbf{n}(t,\rho=1)$. To see that this is indeed so, we consider the configuration $\mathbf{n}(t,\rho)$ as a mapping from two-dimensional disk $(t,\rho)\in \cal{B}_+$ into the two-dimensional sphere $\mathbf{n}\in S^2$. Suppose now that we use another extension $\mathbf{n}'(t,\rho)$ and represent it as a mapping of another disk $\cal{B}_-$ with the same boundary (physical time) into $S^2$. We have
\bea
	W_0[\mathbf{n}]-W_0[\mathbf{n}'] 
	&=&\int_{\cal{B}_+}d^2x\, \frac{1}{8\pi} \epsilon^{\mu\nu}\mathbf n\cdot
	\left[\partial_{\mu}\mathbf n \times\partial_{\nu}\mathbf  n\right]
 \nonumber \\
	&-& \int_{\cal{B}_-}d^2x\, \frac{1}{8\pi} \epsilon^{\mu\nu}\mathbf{n}'\cdot
	\left[\partial_{\mu}\mathbf{n}' \times\partial_{\nu}\mathbf {n}'\right]
 \nonumber \\
    &=& \int_{S^2=\cal{B}_+\cup \cal{B}_-}d^2x\, \frac{1}{8\pi}
    \epsilon^{\mu\nu}\mathbf n\cdot\left[\partial_{\mu}\mathbf n
    \times\partial_{\nu}\mathbf  n\right] = k \,,
 \la{eq:wzvark}
\eea 
where we changed the orientation of $\cal{B}_-$ and considered $\cal{B}_\pm$ as an upper (lower) part of some two-dimensional sphere (see Fig.\ref{fig:WZmv}).
One can recognize the last integral \cite{nakahara2003geometry} as a winding number $k$ of the first sphere ($\cal{B}_+\cup \cal{B}_-$ around the second $\mathbf{n}\in S^2$. This number is always integer proving that $e^{2\pi i W_0}$ does not depend on the particular way of an extension $\mathbf{n}(t,\rho)$.
We notice here that in general the topological term $W_0$ can appear in the action only with the coefficient which is a multiple of $2\pi i$. Otherwise, it depends on the unphysical values of $\mathbf{n}(t,\rho)$ and is not defined. Such a term is called\footnote{It is also often called WZW or Wess-Zumino-Witten or even WZWN or Wess-Zumino-Novikov-Witten term to honor E. Witten\cite{1983-Witten-global,Witten-Quark-1983} and S.P. Novikov\cite{1982-Novikov}.} ``Wess-Zumino term'' or ``WZ term'' by names of Wess and Zumino who discovered a similar term first in the context of four-dimensional quantum field theories\cite{1971-WessZumino}. 
If Wess-Zumino term is present with some coupling constant $g$ so that the weight in partition function is proportional to $e^{2\pi i gW_0}$ we immediately conclude that $g$ must be an integer.
This phenomenon is called ``topological quantization'' of physical constant $g$ and is a very important consequence of Wess-Zumino term.

To obtain the equations of motion from (\ref{cas},\ref{wz0}) we use (\ref{WZvar}) and introduce Lagrange multiplier $\lambda$ to enforce constraint $\mathbf{n}^2=1$. Then we obtain for the variation of the action
\be
 \la{vswz}
    \delta_n\left(S[\mathbf{n}]+\lambda(\mathbf{n}^2-1)\right) 
    = -4\pi S \frac{1}{4\pi}\left[\dot{\mathbf{n}}\times\mathbf{n}\right]
    +S\mathbf{h} + 2\lambda \mathbf{n} = 0\,.
\ee
Vector-multiplying (\ref{vswz}) from the right by $\mathbf{n}$ we arrive at (\ref{eqmcl}). 

In this simplified treatment we just found some classical action which reproduces the classical limit of operator equations of motion (\ref{eqmop}). One can proceed more formally starting with  commutation relations (\ref{crsu2}) and quantum Hamiltonian (\ref{smf}) and derive the classical action (\ref{cas}) using, e.g., coherent states method\cite{fradkin2013field}.
 
The purpose of this exercise was to illustrate that the Wess-Zumino term $W_0$ summarizes at the classical level the commutation relations (\ref{crsu2}). One can also show that reversely the path integral quantization of (\ref{vswz}) produces the commutation relations (\ref{crsu2}).

\subsection{Fermionic model}

In this section, we use a very simple quantum mechanical example to show how topological terms are generated when one passes from microscopic theory to an effective description. Generally, in condensed matter physics we are dealing with some system of electrons interacting with each other as well as with other degrees of freedom such as a lattice. Let us assume that at some low energy scale we reduced our problem to fermions interacting to a bosonic field. The bosonic field may originate both from the collective behavior of electrons, e.g., magnetization or superconducting order parameter, and from independent degrees of freedom, e.g., from the vibrations of the lattice. For our illustrative example we consider \cite{1986-Stone}
\be
    S = \int dt\, \psi^\dagger\left[i\partial_t + m \mathbf{n}\cdot\bm{\sigma}\right]\psi,
 \la{ex1wz}
\ee
where $m$ is a coupling constant, $\psi =(\psi_1,\psi_2)^t$ is a spinor, and $\bm{\sigma}$ is a triplet of Pauli matrices. In this case, fermions are represented by just one spinor and the bosonic field by a single unit vector $\mathbf{n} =(n_1,n_2,n_3)$, $\mathbf{n}\in S^2$. The latter means that $\mathbf{n}$ takes its values on a two-dimensional sphere, i.e., $\mathbf{n}^2=1$.
This model can originate, e.g., from electrons interacting with a localized magnetic moment. Then coupling constant $m>0$ corresponds to a Hund's coupling between electrons (one electron for simplicity) and the direction $\mathbf{n}$ of a localized moment. Notice, that a more complete theory must have the bare action of a moment $\mathbf{n}$ added to a (\ref{ex1wz}). We, however, are interested only in the action of $\mathbf{n}$ induced by an interaction with fermions.

For future convenience, we will use a Euclidean formulation here and in the rest of the paper. It can be obtained by ``Wick rotation'' $t\to it$. A Euclidean action obtained from  (\ref{ex1wz}) is
\be
    S_{E} = \int dt\, \psi^\dagger\left[\partial_t -  m \mathbf{n}\cdot\bm{\sigma}\right]\psi.
 \la{ex1wzE}
\ee

\subsubsection{Effective action by chiral rotation trick}

We consider partition function 
\be
    Z = \int{\cal D}\psi\,{\cal D}\bar{\psi}\,{\cal D}\mathbf n \, e^{-S_E}
      = \int{\cal D}\mathbf n \, e^{-S_{eff}},
    \la{eq:sqpiz0}
\ee
where the last equality is a definition of an effective action 
\be
    S_{eff} = -\ln \int{\cal D}\psi\,{\cal D}\bar{\psi}\, e^{-S_E}
      = -\ln \det D,
    \la{eq:seff}
\ee
where we defined an operator $D \equiv \partial_t -  m \mathbf{n}\cdot\bm{\sigma}$.
To calculate the logarithm of the determinant we use ``chiral rotation''. Namely, we introduce the matrix field $U(t)\in SU(2)$ such that $U^\dagger \mathbf{n}\cdot\bm{\sigma} U = \sigma^3$ so that 
\be
    \tilde D = U^\dagger D U = \partial_t -i \hat{a} -m\sigma^3=G_0^{-1}-i\hat{a},
\ee
with 
\be
    \hat{a} \equiv U^\dagger i\partial_t U
 \la{adef}
\ee
and 
\be
	G_0 = (\partial_t -m \sigma^3)^{-1}\,.
\ee
Then we write\footnote{Notice that the second equality in (\ref{eq:seffRot}) is the common source of miscalculated topological terms. Quantum anomalies might be present making chiral rotation technique inapplicable. In this case this is a legitimate procedure because of the absence of so-called global anomalies \cite{1982-Witten-SU2,abanov2000hopf}.}
\be
    S_{eff} = -\ln \det D = -\ln \det \tilde D = -\Tr \ln \tilde D.  
    \la{eq:seffRot}
\ee
Let us now write $\tilde D = G_0^{-1}(1-G_0 i\hat{a})$ and expand
\bea
    S_{eff} =  -\Tr \ln \tilde D &=& \Tr \left[\ln G_0 +G_0 i\hat{a} 
            +\frac{1}{2}(G_0 i\hat{a})^2 +\ldots \right]
 \nonumber \\
            &=&  S^{(0)} + S^{(1)} + S^{(2)} + \ldots.  
    \la{eq:seffRotExp}
\eea
The expansion 
(\ref{eq:seffRotExp}) has the following diagrammatic representation 
\be
     S_{eff} = const + \parbox{20mm}{\includegraphics[width=1.5cm]{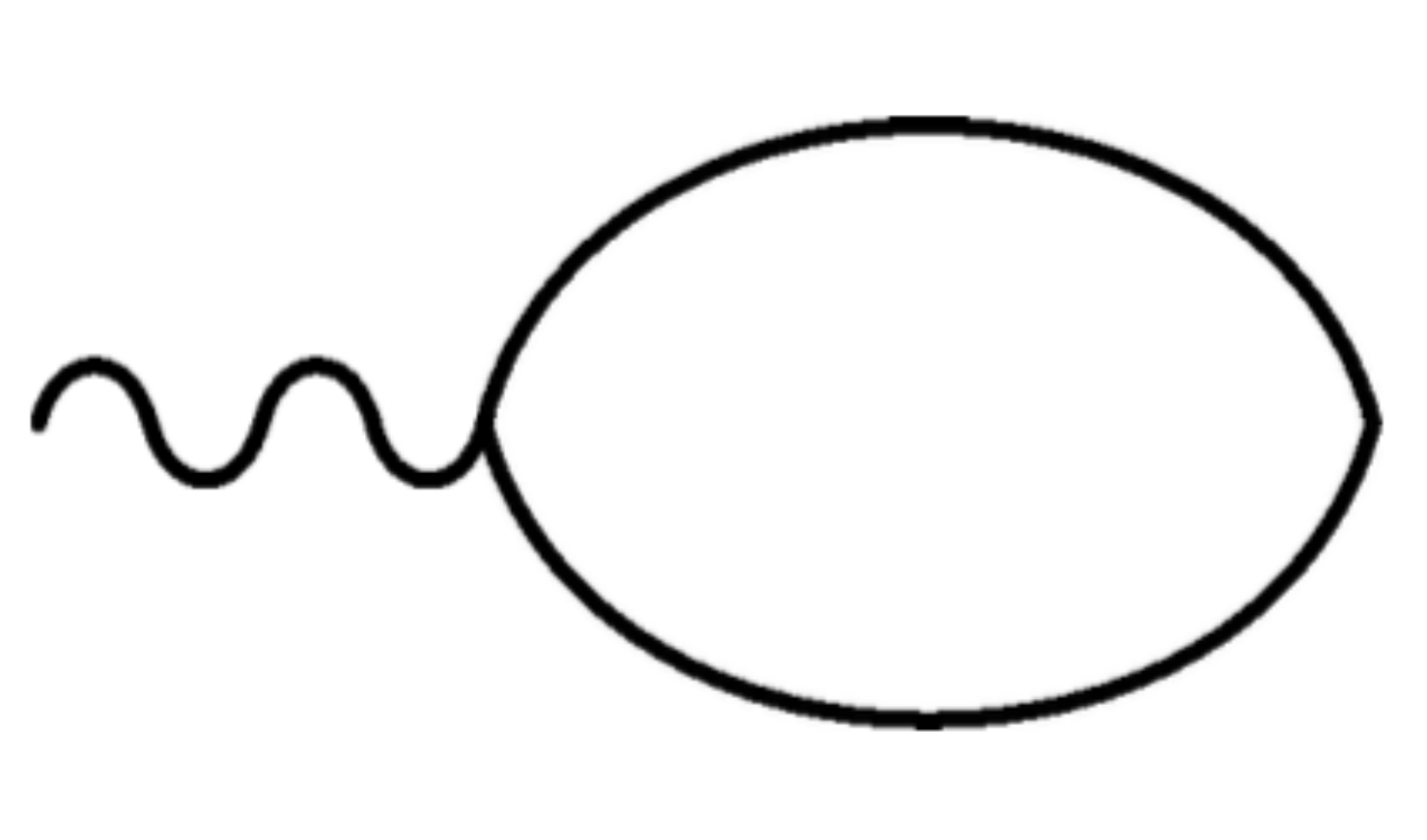}} 
     + \frac{1}{2} \parbox{20mm}{\includegraphics[width=2cm]{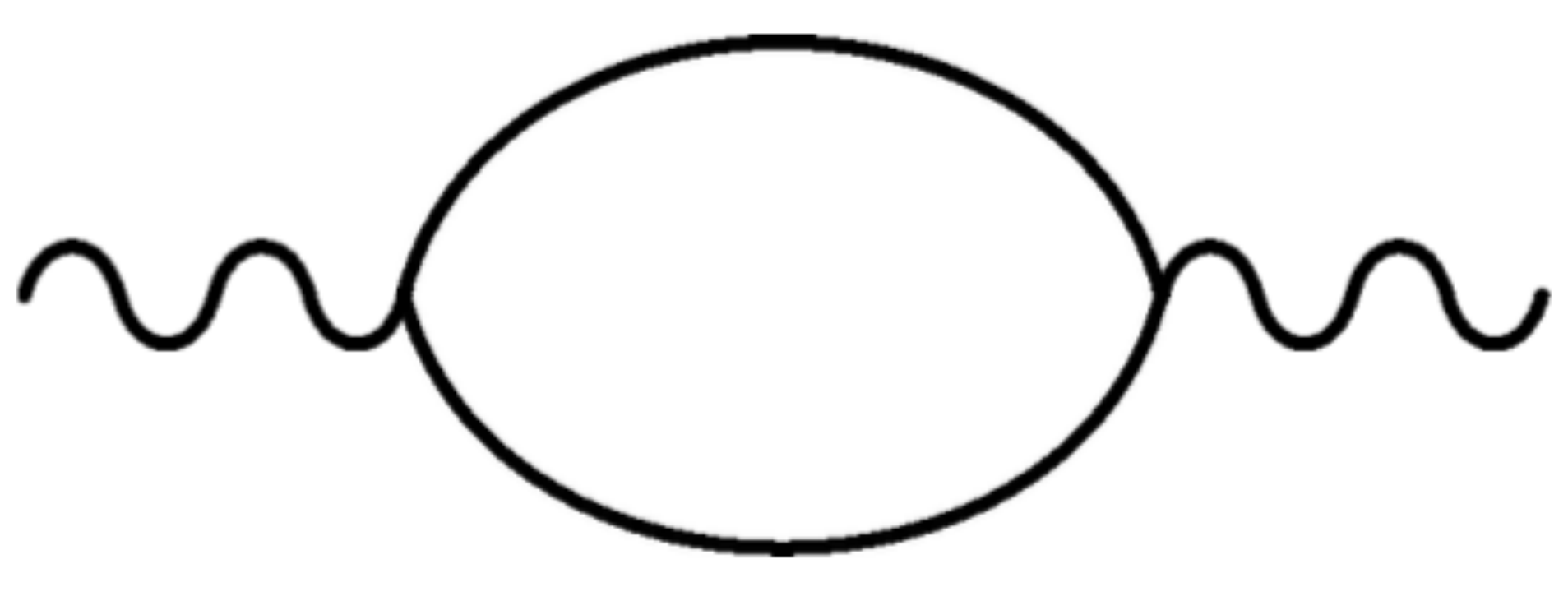}} +\ldots.
\ee
The zeroth order term $S^{(0)}$ is an (infinite) constant which does not depend on $\hat{a}$. The first order term is given by 
\be
    S^{(1)} = \parbox{20mm}{\includegraphics[width=1.5cm]{Figures/feyn-diag1}} 
    = \Tr \left[G_0 i\hat{a} \right]
    = \int \frac{d\omega}{2\pi}\, 
    \tr \left[\frac{1}{-i\omega - m\sigma^3}i\hat{a}_{\omega=0}\right].
\ee
Here $\hat{a}_{\omega=0} = \int dt\, \hat{a}(t)$ and $\tr$ is taken over sigma-matrices. Closing the integral in an upper complex $\omega$-plane we obtain
\be
    S^{(1)} = \int \frac{d\omega}{2\pi}\, 
    \tr \left[\frac{1}{-i\omega - m\sigma^3}i\hat{a}_{\omega=0}\right]
    = -ia_{\omega=0}^3 = -i\int dt\, a^3(t), 
 \la{seff1}
\ee
where only the term containing $a^3$ ($\hat{a} = a^k\sigma^k$, $k=1,2,3$) does not vanish when trace over Pauli matrices is taken.\footnote{We notice that the $\omega$-integral in (\ref{seff1}) is formally diverging. However, being regularized, it becomes the number of fermions in the system. In our model we have exactly one fermion and the regularization procedure in this case is just the closing of the contour of an integration in an upper complex plane.}

We may proceed and obtain for the second term of an expansion
\bea
    S^{(2)} &=& \frac{1}{2} \parbox{20mm}{\includegraphics[width=2cm]{Figures/feyn-diag2} }
    = \frac{1}{2} \Tr \left[G_0 i\hat{a} G_0 i\hat{a}\right]
 \nonumber \\
     &=&  \frac{1}{2}\int \frac{d\Omega}{2\pi}\,\int \frac{d\omega}{2\pi}\, 
    \tr \left[\frac{1}{-i\omega - m\sigma^3}i\hat{a}_{-\Omega}
    \frac{1}{-i(\omega+\Omega) - m\sigma^3}i\hat{a}_{\Omega}\right] 
 \nonumber \\
    &=& \frac{1}{8 m}\int \frac{d\Omega}{2\pi}\,
    \tr \left[\hat{a}_{-\Omega}\hat{a}_{\Omega}
    -\sigma^3 \hat{a}_{-\Omega} \sigma^3 \hat{a}_{\Omega}\right] 
    +o\left(\frac{1}{m}\right)
 \nonumber \\
    &=& \frac{1}{2 m} \int dt\,\left[(a^1)^2+(a^2)^2\right]
    +o\left(\frac{1}{m}\right).
\eea
We neglected here the terms of higher order in $1/m$. Therefore, for an effective action we obtain up to the terms of the order of $1/m$ and omitting constant
\be
    S_{eff} = S^{(1)} + S^{(2)} =  -i\int dt\, a^3(t) 
    + \frac{1}{2 m} \int dt\,\left[(a^1)^2+(a^2)^2\right].
 \la{seff12}
\ee
The effective action (\ref{seff12}) is expressed in terms of an auxiliary gauge field $\hat{a}$. However, one should be able to re-express it in terms of physical variable $\mathbf{n}$ as it was defined by  (\ref{eq:seff}) which contains only $\mathbf{n}$. Let us start with the second term. Using an explicit relation $\mathbf{n}\cdot\bm{\sigma}=U\sigma^3 U^\dagger$ and the definition (\ref{adef}) one can easily check that $(\partial_t\mathbf{n})^2 = 4\left[(a^1)^2+(a^2)^2\right]$ and  the last term of (\ref{seff12}) indeed can be expressed in terms of $\mathbf{n}$ as
\be
    S^{(2)} = \frac{1}{2 m} \int dt\,\left[(a^1)^2+(a^2)^2\right]
    =\frac{1}{8 m} \int dt\,(\partial_t\mathbf{n})^2.
 \la{seff2n}
\ee
Obtaining $S^{(1)}$ is a bit more subtle. The gauge field is defined as (\ref{adef}) with matrix $U$ defined implicitly by $\mathbf{n}\cdot\bm{\sigma} = U\sigma^3 U^\dagger$. One can see from the latter expression that the definition of $U$ is ambiguous. Indeed, one can make a ``gauge transformation''
\be
    U\to U e^{i\sigma^3\psi}
 \la{gt}
\ee
with $\psi(t)$ any function of $t$ without changing $\mathbf{n}$.
Under this transformation the gauge field is transformed as $\hat{a} \to e^{-i\sigma^3\psi} \hat{a} e^{i\sigma^3\psi} -\sigma^3 \partial_t\psi$, or
\bea
    a^3 & \to & a^3 - \partial_t \psi, \\
    a^1 & \to & a^1\cos 2\psi - a^2 \sin 2\psi, \\ 
    a^2 & \to & a^1\sin 2\psi + a^2 \cos 2\psi.
\eea
Therefore, $S^{(1)} \to S^{(1)} +i \int dt\, \partial_t \psi$ and we notice that $S^{(1)}$ transforms non-trivially\footnote{Notice that (\ref{seff2n}) does not transform under this  gauge transformation.} under the change of $U$ and therefore, can not be expressed as a simple time integral over the function which depends on $\mathbf{n}$ only. One might question the validity of our derivation because it seems that $S^{(1)}$ defined by (\ref{seff1}) is not invariant under the transformation (\ref{gt}) but we notice that $S^{(1)}$ changes only by the integral of a full time derivative. Moreover, if we require periodicity in time, i.e., time changes from $0$ to $\beta$ and $\psi(\beta) = \psi(0) + 2\pi n$ with an integer $n$, then
$S^{(1)}\to S^{(1)}+2\pi i n$ and ``Boltzmann'' weight $e^{-S^{(1)}}$ is invariant under (\ref{gt}). Therefore, the contribution to the partition function from the $S^{(1)}$ term depends only on the physical variable $\mathbf{n}$. To understand what is going on let us calculate $S^{(1)}$ explicitly. We parametrize $\mathbf{n} = (\sin\theta\cos\phi, \sin\theta\sin\phi, \cos\theta)$. Then, the most general choice of $U$ is
$$ 
U =\left( \begin{array}{cc}
          \cos\frac{\theta}{2} & e^{-i\phi} \sin\frac{\theta}{2} \\
          e^{i\phi} \sin\frac{\theta}{2} & \cos\frac{\theta}{2}
          \end{array} \right) 
   e^{i\sigma^3\psi},
$$
where $\psi(t)$ is an arbitrary function of $t$ with $\psi(\beta) = \psi(0) +2\pi n$. It is straightforward to calculate
$$
a_3 = -\frac{1-\cos\theta}{2}\partial_t\phi -\partial_t\psi.
$$
We see that the last term can be discarded by reasons given above, and we have
\be
    S^{(1)}  =  2\pi i W_{0},
 \la{wz0eff}
\ee
with $W_{0}$ defined in (\ref{seff1loc}).

Combining (\ref{seff2n}) and (\ref{wz0eff}) together we obtain
\be
    S_{eff} = 2\pi i W_{0} + \frac{1}{8 m} \int dt\,(\partial_t\mathbf{n})^2 
    +o\left(\frac{1}{m}\right).
 \la{seff0sum}
\ee

In the case where $N$ species of fermions coupled to the same $\mathbf{n}$ field are present, one obtains an overall factor $N$ in effective action, i.e., $S_{eff}\to N S_{eff}$. 

\subsubsection{Topological term}

Let us notice that the first term $2\pi iW_0$ of gradient expansion (\ref{seff0sum}) is very different from, say, the second one in the following respects
\begin{enumerate}
\item It is imaginary.
\item It does not depend on the value of the mass parameter $m$.\footnote{If we allow $m$ to be negative this term becomes $2\pi i (\sgn m) W_0$ and depends only on the sign of $m$, not its magnitude.}
\item It does not change under reparameterization of time $t\to f(t)$. In particular, it is scale invariant and does not change when $t\to \lambda t$.
\end{enumerate}
The property 3 makes it natural to call the term $2\pi i\,W_0$ topological as it does not depend on time scales but only on the trajectory of $\mathbf{n}(t)$. In fact, properties 1, 2 are consequences of 3 and are the general properties of all topological terms.

\subsubsection{Path integral representation of quantum spin}

Before going to the next section let us consider some application of derived topological term. We generalize our model slightly so that in (\ref{ex1wzE}) $\psi$ denotes $N$ species of fermions which are all coupled to the same bosonic field $\mathbf{n}$. We consider the special limit $m\to \infty$ of the model (\ref{ex1wzE}). The Hamiltonian of the model $-m\mathbf{n}\cdot \psi^\dagger\bm{\sigma}\psi$ in this limit forces all spins of $\psi$ particles to be aligned along $\mathbf{n}$. Therefore, we expect that, in this limit, after an integration over fermions, we will obtain an effective action of a quantum spin $S=N/2$ written in terms of the direction $\mathbf{n}$ of its quantization axis. Multiplying (\ref{seff0sum}) by the number of fermion species $N$, and taking limit $m\to\infty$, we obtain
\begin{equation}
    S = 2\pi i N W_{0}.
    \label{spin12}
\end{equation}
One can show that upon quantization\footnote{The easy way to show that we are dealing with the  spin is to add coupling to an external magnetic field $-\int dt\,S\mathbf{h}\cdot\mathbf{n}$ to (\ref{spin12}) and write down the classical equation of motion for $\mathbf{n}$ using (\ref{WZvar}) and constraint $\mathbf{n}^2=1$. We obtain $\dot{\mathbf{n}} = [\mathbf{n}\times\mathbf{h}]$  where we assumed $S=N/2$ and changed to the real time $t\to it$. The obtained equation is indeed the classical equation of spin precession.} the components of $\mathbf{n}$ become the components of the quantum spin $S=N/2$ so that $n_a\to \hat{S}_a/S$.
The action (\ref{spin12}) is explicitly $SU(2)$ invariant and  is well-defined for integer $2S=N$, i.e. spin can only be integer or half-integer. Therefore, in path integral formulation the quantization of spin is a consequence of the Wess-Zumino term in the action of the spin.

\subsection{Derivation of a WZ term from fermionic model without chiral rotation}
\la{sec:wzferm}

Here we give an alternative derivation of an effective action (\ref{seff0sum}) from (\ref{eq:seff}) which does not use chiral rotation trick. \cite{AbanovWiegmann-2000}

Consider 
\begin{equation}
    Z = \int{\cal D}\psi\,{\cal D}\bar{\psi}\,{\cal D}\mathbf n \, e^{-S},
    \label{eq:sqpiz}
\end{equation}
where 
\begin{equation}
    S= \int_{0}^{T}dt\, 
    \bar\psi\left(i\partial_{t}-im \mathbf n\cdot \bm\tau\right)\psi.
 \label{eq:sqpis}
\end{equation}
Here $\psi=(\psi_{1},\psi_{2})$ is a Grassmann spinor representing 
spin $1/2$ fermion, $\bm\tau$ are the Pauli matrices acting on spinor 
indices of $\psi$, and ${\mathbf n}^{2}=1$ is a unit, three-component 
vector coupled to 
the spin of fermion $\bar\psi\bm\tau\psi$ with the coupling constant 
$m$.
Integrating out fermions in (\ref{eq:sqpiz}) we obtain
\begin{equation}
    Z = \int{\cal D}\mathbf n \, e^{-S_{\rm eff}(\mathbf n)},
    \label{eq:sqpizeff}
\end{equation}
with
\begin{equation}
    S_{\rm eff}(\mathbf n)
    = -\ln \det \left(i\partial_{t}-im \mathbf n\cdot \bm\tau\right).
    \label{eq:sqpiseff}
\end{equation}
Let us denote $D=i\partial_{t}-im \mathbf n\cdot \bm\tau$ and 
$D^{\dagger}=i\partial_{t}+im \mathbf n\cdot \bm\tau$. We calculate the 
variation of the effective action
\begin{eqnarray}
    \delta S_{\rm eff} &=& -\mbox{Tr}\,\left\{\delta D D^{-1}\right\} 
    =-\mbox{Tr}\,\left\{\delta D D^{\dagger}(DD^{\dagger})^{-1}\right\} 
 \nonumber \\
    &=&  im \mbox{Tr}\,\left\{\delta \mathbf n\cdot\bm\tau(i\partial_{t}
    +im\mathbf  n\cdot\bm\tau)
    (-\partial_{t}^{2}+m^{2}-m\dot{\mathbf n}\cdot\bm\tau)^{-1} \right\}
\end{eqnarray}
Expanding the fraction in $\dot{\mathbf n}$, calculating the trace, and 
keeping only lowest orders in $\dot{\mathbf n}/m$, we obtain:
\begin{equation}
    \delta S_{\rm eff}  = \int dt\, \left\{ 
    \frac{1}{4m}\delta\dot{\mathbf  n}\cdot\dot{\mathbf n}
    -\frac{i}{2}\delta\mathbf n\cdot\left[\mathbf n\times\dot{\mathbf 
    n}\right]\right\}.
 \label{eq:wz0var}
\end{equation}
Restoring  the effective action from its variation we have:
\begin{equation}
    S_{\rm eff}  = \int_{0}^{T} dt\, 
    \frac{1}{8m}\dot{\mathbf  n}^{2} 
    -2\pi i W_{0},
 \label{eq:wz0eff}
\end{equation}
where the {\em Wess-Zumino action} 
\begin{equation}
    W_{0}=\int_{0}^{1}d\rho\, \int_{0}^{T} dt\, \frac{1}{8\pi}
    \epsilon^{\mu\nu}\mathbf n\cdot\left[\partial_{\mu}\mathbf n
    \times\partial_{\nu}\mathbf  n\right].
 \label{eq:wz0}
\end{equation}
Here $\rho$ is an auxiliary coordinate $\rho \in\left[0,1\right]$. 
Also, the $\mathbf n$-field is extended to $\mathbf n(t,\rho)$ in such a way that 
$\mathbf n(t,0)=(0,0,1)$ and $\mathbf n(t,1)=\mathbf n(t)$. Indices $\mu,\nu$ 
take values $t,\rho$.
\vspace{0.5cm}

The Wess-Zumino action (\ref{eq:wz0}) has a very special property. 
Although it is defined as an integral over two-dimensional disk 
parameterized by $\rho$ and $t$ its variation depends only on the 
values of $\mathbf n$ on the boundary of the disk---physical time.

\subsection{Quantum spin as a particle moving in the field of Dirac 
monopole}

Let us think of the action (\ref{eq:wz0eff}) as of the action of a charged particle moving on a surface of two-dimensional sphere with unit radius so that $\mathbf n$ is a position of particle on the sphere. Then the first term of (\ref{eq:wz0eff}) is a conventional kinetic energy of the particle. The second term should then be interpreted as a phase picked by particle 
moving in the field a magnetic charge $2S$ (Dirac monopole) placed in the center of the sphere.

One could have started with the problem of particle of the mass $m$ 
moving in the field of magnetic monopole of charge $2S$. Then the 
ground state is $2S$-degenerate and is separated by the gap $\sim 
1/m$ from the rest of the spectrum. In the limit $m\to 0$ only the 
ground state is left and we obtain a quantum spin problem in an approach 
analogous to the plane rotator from Sec.\ref{sec:QD}.

\subsection{Reduction of a WZ term to a theta-term}
\la{sec:redwztheta}

Let us consider the value of (\ref{seff1loc}) assuming that the polar angle is kept constant at $\theta(\tau)=\theta_{0}$. Then (\ref{seff1loc}) becomes
\be 
	W_{0}=\frac{1-\cos\theta_{0}}{2}\int_{0}^{\beta}\frac{d\tau}{2\pi}\,\partial_{\tau}\phi ,
 \la{w0theta}
\ee
and we recognize (\ref{wz0eff}) with (\ref{w0theta}) as the theta-term (\ref{ringtop}) corresponding to the particle on a ring with the flux through the ring given by
\be
	A = \frac{1-\cos\theta_{0}}{2}.
\ee
In particular, $\theta_{0}=\pi/2$ the topological term in the action of a particle on a ring in
magnetic field  $A=1/2$.

\subsection{Properties of WZ terms}

WZ terms 
\begin{enumerate}
	\item do not depend on the metric of spacetime
	\item are imaginary in Euclidean formulation
	\item do not contribute to stress-energy tensor (and to Hamiltonian).
	\item do not depend on $m$ -- the scale, below which an effective action is valid (but do depend on $\sgn(m)$)
	\item are antisymmetric in derivatives with respect to different space-time coordinates (contain $\epsilon^{\mu\nu\lambda\ldots}$)
	\item are written as integrals of (D+1)-forms over auxiliary $(D+1)$-dimensional space - disk $D^{D+1}$ such that $\partial D^{D+1}=S^{D}$ - compactified space-time
	\item are multi-valued functionals. Multi-valuedness results in quantization of coupling constants (coefficients in front of WZ terms)
	\item do change equations of motion by changing commutation relation between fields (Poisson's brackets), not by changing Hamiltonian
	\item might lead to massless excitations with ``half-integer spin'' (see Sec.~\ref{sec:bound-states-haldane})
	\item describe boundary theories of models with $\theta$-terms (see Sec.~\ref{sec:bound-states-haldane})
	\item being combined (see the spin chains Sec.~\ref{sec:cpiQAFM}) produce $\theta$-terms
	\item can be calculated by gradient expansion of the variation of fermionic determinants
	\item produce $\theta$ terms as a reduction of target space (see the Sec.~\ref{sec:redwztheta}) \cite{AbanovWiegmann-2000}
\end{enumerate}
Among the listed  properties the first five 1-5 are the properties of all topological terms while the others are more specific to WZ terms. 

\subsection{Exercises}


The exercises (\ref{ex:wz01}-\ref{ex:wz01ferm}) were solved in the main text. Try to solve them independently and test your understanding by solving exercises (\ref{ex:det2D}-\ref{ex:chirrot}).

{\small
\prn{ex:wz01}
\subsubsection*{Exercise \ref{ex:wz01}: WZ term in $0+1$, preliminaries}

Consider a three-dimensional unit vector field $\mathbf{n}(x,y)$ ($\mathbf{n}\in S^{2}$)  defined on a two-dimensional disk $D$. Define
\begin{equation}
 \label{wzw0}
    W_{0} =  \int_{D} d^{2}x\, \frac{1}{8\pi}
    \epsilon^{\mu\nu}\mathbf{n}\cdot[\partial_{\mu}\mathbf{n}\times\partial_{\nu}\mathbf{n}]
    = \int_{D} \frac{1}{16\pi i }\tr [\hat n d\hat n d\hat n],
\end{equation}
where the latter expression is written in terms of differential forms and $\hat n =\mathbf{n}\cdot\bm{\sigma}$. 

a) Calculate the variation of $W_{0}$ with respect to $\mathbf{n}$. Show that the integral becomes the integral over disk $D$ of the complete divergence (of the exact form). 

b) Parametrize the boundary $\partial D$ of the disk by parameter $t$, apply Gauss-Stokes theorem and express the result of the variation using only the values of $\mathbf{n}(t)$ at the boundary.

We showed that the variation of $W$ depends only on the boundary (i.e. physical) values of $\mathbf{n}$-field. See (\ref{WZvar}) for the answer. 

\prn{ex:wz01def}
\subsubsection*{Exercise \ref{ex:wz01def}: WZ term in $0+1$, definition}

Assume that we are given the time evolution of $\mathbf{n}(t)$ field ($\mathbf{n}\in S^{2}$). We also assume that time can be compactified, i.e. $\mathbf{n}(t=\beta)=\mathbf{n}(t=0)$. Consider the two-dimensional disk $D$ which boundary $\partial D$ is parametrized by time $t\in[0,\beta]$. The WZ term is defined by 
\be
	S_{WZ} =i 4\pi S \, W_{0}[\mathbf{n}],
 \la{eq:wz07}
\ee
where $S$ is some constant, $W_{0}$ is given by Eq. (\ref{wzw0}), and  $\mathbf{n}(x,y)$ is some arbitrary smooth extension of $\mathbf{n}(t)$ from the boundary to an interior of the disk.

Let us show that the WZ term is well defined and (almost) does not depend on the extension of $\mathbf{n}(t)$ to the interior of $D$.

Consider two different extensions $\mathbf{n}^{(1)}(x,y)$ and $\mathbf{n}^{(2)}(x,y)$ of the same $\mathbf{n}(t)$ and corresponding values $W_{0}^{(1)}$ and $W_{0}^{(2)}$ of the functional $W_{0}$. Show that the difference $W_{0}^{(1)}-W_{}^{(2)}$ is an integer number - the degree $Q$ of mapping $S^{2}\to S^{2}$. The second $S^{2}$ here is a target space of $\mathbf{n}$. How did the first $S^{2}$ appear?

We see that $S_{WZ}[\mathbf{n}(t)]$ is a multi-valued functional which depends on the extension of $\mathbf{n}$ to the disk $D$. However, the weight in partition function is given by $e^{-S_{WZ}}$ and can be made single-valued functional if the coupling constant $S$ is ``quantized''. Namely, if $2S\in \mathbf{Z}$ ($S$ - half-integer number) the $e^{-S_{WZ}}$ is a well-defined single-valued functional.

For the answer see (\ref{eq:wzvark}).

\prn{ex:wz01spin}
\subsubsection*{Exercise \ref{ex:wz01spin}: WZ in $0+1$, spin precession}

Let us consider the quantum-mechanical action of the unit vector $\mathbf{n}(t)$ with the (Euclidean) action
\be
	S_{h} = S_{WZ}[\mathbf{n}(t)] -S \int dt\, \mathbf{h}\cdot\mathbf{n}(t),
 \label{sh}
\ee
where $S_{WZ}$ is given by (\ref{eq:wz07}) and $\mathbf{h}$ is some constant three-component vector (magnetic field). 

Find the classical equation of motion for $\mathbf{n}(t)$ from the variational principle $\delta S_{h}=0$. Remember that one has a constraint $\mathbf{n}^{2}=1$ which can be taken into account using, e.g., Lagrange multiplier trick.

The obtained expression is the equation of spin precession and $S_{WZ}$ is a proper, explicitly $SU(2)$ invariant action for the free spin $S$.

For the answer see (\ref{vswz},\ref{eqmcl}).

\prn{ex:wz01q}
\subsubsection*{Exercise \ref{ex:wz01q}: WZ in $0+1$, quantization}

Show that the classical equations of motion obtained from $S_{h}$ correspond to Heisenberg equations (in real time) $\partial_{t}\hat{\mathbf{S}} =i \left[H,\hat{\mathbf{S}}\right]$ for the quantum spin operator $\hat{\mathbf{S}}$
\be
	\left[S^{a},S^{b}\right]=i\epsilon^{abc}S^{c}
 \la{crspin}
\ee 
obtained from the Hamiltonian of a spin in magnetic field
\be
	H = -\mathbf{h}\cdot\hat{\mathbf{S}}.
\ee 

Obtain the commutation relations of quantum spin (\ref{crspin}) from the topological part $S_{WZ}$. Notice that this topological action is linear in time derivative and, therefore, does not contribute to the Hamiltonian. Nevertheless, it defines commutation relations between components of the spin operator. 

\textit{Hint}: You can either use local coordinate representation of the unit vector in terms of spherical angles $\mathbf{n} = (\cos\phi\sin\theta, \sin\phi\sin\theta, \cos\theta)$ or use the general formalism of obtaining Poisson bracket from the symplectic form given in $S_{WZ}$.

\prn{ex:wz01red}
\subsubsection*{Exercise \ref{ex:wz01red}: Reduction of the WZ-term to the theta-term in $0+1$}

Let us assume that the field $\mathbf{n}(t)$ is constrained so that it takes values on a circle given in spherical coordinates by $\theta =\theta_{0}=const$. Find the value of the topological term $S_{WZ}$ on such configurations (notice that this constraint is not applicable in the interior of the disk $D$, only at its physical boundary). Show that the obtained topological term is a theta-term in $0+1$ corresponding to $S^{1}\to S^{1}$. 

What is the value of the coefficient in front of that topological term? What is the value of corresponding ``magnetic flux'' through a ring? For $S=1/2$ which reduction (value of $\theta_{0}$) corresponds to the half of the flux quantum?

For the answer see Sec.~\ref{sec:redwztheta}.

\prn{ex:wz01ferm}
\subsubsection*{Exercise \ref{ex:wz01ferm}: WZ in $0+1$, derivation from fermions}

Consider a Euclidean action of a fermion coupled to a unit vector
\be
	S_{E} = \int d\tau\, \psi^{\dagger} D \psi,
\ee
where
\be
	D = \partial_{\tau} -m\mathbf{n}\cdot\bm{\tau},
\ee
with $\mathbf{n}\in S^{2}$ and $\bm\tau$ the vector of Pauli matrices. We obtain an effective action for $\mathbf{n}$ induced by fermions as 
\be
	e^{-S_{eff}} = \int D\psi D\psi^{\dagger} e^{-S_{E}} = \Det D,
\ee
or 
\be
	S_{eff} = -\log \Det D = -\Tr \log D.
\ee
We calculate the variation of $S_{eff}$ with respect to $\mathbf{n}$ as 
\bea
	\delta S_{eff} &=& -\Tr \delta D D^{-1} = -\Tr \delta D D^{\dagger} (DD^{\dagger})^{-1},
 \la{seffdddagger}
\eea
where $D^{\dagger} = -\partial_{\tau} -m\mathbf{n}\cdot\bm{\tau}$. We have
\bea
	DD^{\dagger} &=& -\partial_{\tau}^{2} +m^{2} -m\dot{\mathbf{n}}\cdot \bm{\tau}
	=G_{0}^{-1} -m\dot{\mathbf{n}}\cdot \bm{\tau}.
\eea

Expand (\ref{seffdddagger}) in $1/m$ up to the term $m^{0}$ and calculate functional traces. Show that the term of the order $m^{0}$ is a variation of the WZ term in 0+1 dimensions. Restore $S_{eff}$ from its variation. What is the coefficient in front of the WZ term? To what value of spin does it correspond?

For the answer see Sec.~\ref{sec:wzferm}.

\prn{ex:det2D}
\subsubsection*{Exercise \ref{ex:det2D}: Fermionic determinant in two dimensions}

Let us consider two-dimensional fermions coupled to a phase field $\phi(x)$ ($\phi\equiv \phi+2\pi$). The Euclidean Lagrangian is given by
\begin{equation}
	{\cal L}_{2} = \bar\psi\left[i\gamma^{\mu}(\partial_{\mu}-iA_{\mu})+im e^{i\gamma^{5}\phi}
	\right]\psi,
\end{equation}
where $\mu=1,2$ is a spacetime index, $\gamma^{1,2,5}$ is a triplet of Pauli matrices, and $A_{\mu}$ is an external gauge field probing fermionic currents.

We assume that the bosonic field $\phi$ changes slowly on the scale of the ``mass'' $m$. Then one can integrate out fermionic degrees of freedom and obtain an induced effective action for the $\phi$-field as a functional determinant.
\begin{eqnarray}
	S_{eff} &=&  -\log\mbox{Det}\, D,
 \label{slndphi}\\
 	D &=& i\gamma^{\mu}(\partial_{\mu}-iA_{\mu})+im\, e^{i\gamma^{5}\phi}.
 \label{d2s1}
\end{eqnarray}

We calculate the effective action using the gradient expansion method. Namely, we calculate the variation of (\ref{slndphi}) with respect to the $\phi$ and $A$-fields and use
\begin{equation}
	\delta S_{eff} = -\delta \log\mbox{Det}\, D
	=-\mbox{Tr}\, \delta\log D = -\mbox{Tr}\, \delta D\, D^{-1}
	=-\mbox{Tr}\, \delta D\, D^{\dagger} (DD^{\dagger})^{-1}.
\end{equation}

a) Calculate $DD^{\dagger}$ for (\ref{d2s1}). Observe that this object depends only on gradients of $\phi$-field.

b) Expand $(DD^{\dagger})^{-1}$ in those gradients. This will be the expansion in $1/m$. (It is convenient to introduce notation $G_{0}^{-1}= -\partial_{\mu}^{2}+m^{2}$).

c) Calculate functional traces of the terms up to the order of $m^{0}$. Use the plane wave basis to calculate the trace $\mbox{Tr}\,(\hat{X}) \to \int d^{2}x\,\int \frac{d^{2}p}{(2\pi)^{2}} e^{-i\mathbf{p}\cdot\mathbf{x}}\hat{X} e^{i\mathbf{p}\cdot\mathbf{x}}$.

d) Identify the variation of the topological term in the obtained expression. It contains the antisymmetric tensor $\epsilon^{\mu\nu}$ and is proportional to $\mbox{sgn}\,(m)$. 

e) Remove the variation from the obtained expression and find $S_{eff}$ up to the $m^{0}$ order. 

f) Which terms of the obtained action are topological? Can you write them in terms of differential forms?

For the answer see Ref.~\cite{AbanovWiegmann-2000}.

\prn{ex:chirrot}
\subsubsection*{Exercise \ref{ex:chirrot}: ``Dangers'' of chiral rotation}

Try to calculate the determinant of the previous exercise using ``chiral rotation trick''. Namely, consider chiral rotation $\psi \to e^{-i\gamma^{5}\phi/2}\psi$. Then $\psi^{\dagger}\to \psi^{\dagger}e^{i\gamma^{5}\phi/2}$ and $\bar\psi\to \bar\psi e^{-i\gamma^{5}\phi/2}$. Use the identity $\gamma^{\mu}\gamma^{5}=-i\epsilon^{\mu\nu}\gamma^{\nu}$ and anti-commutativity of Pauli matrices to show that the operator $D(A_{\mu},\phi)$ transforms into 
$$
	\tilde{D}(A_{\mu},\phi) = e^{-i\gamma^{5}\phi/2}D(A_{\mu},\phi) e^{-i\gamma^{5}\phi/2}
	=D(A_{\mu}+\frac{i}{2}\epsilon^{\mu\nu}\partial_{\nu}\phi, 0)=D(\tilde{A}_{\mu},0).
$$
Try to calculate $\log\Det \tilde{D}=\log\Det D(\tilde{A},0)$ using expansion in $\tilde{A}$. 
You will see that the result does not match the effective action obtained in the previous exercise. Why? What one should add to the chiral rotation trick to make the correct calculation?

Answer: The Jacobian of the change of variables corresponding to the chiral rotation. See Refs.~\cite{AbanovWiegmann-2000} and \cite{fujikawa2004path}.

}

\section{Spin chains.}

Here we study how topological terms appear in effective theories for quantum spin chains. We emphasize an interplay between different types of topological terms and the effects of topological terms on field dynamics. In addition to original papers, the useful references for this section include: \cite{PolyakovBook-1987,affleck1988field,fradkin2013field}.

Let us start with the model of quantum magnet
\be
	H = \sum_{<kj>}J_{kj}\mathbf{S}_{k}\cdot \mathbf{S}_{j} -\sum_{j}\mathbf{h}_{j}\cdot\mathbf{S}_{j}\,.
 \la{eq:Hmag}
\ee
Here the summation is taken over the sites $k,j$ of some $d$-dimensional lattice, $J_{kj}$ are exchange integrals and $\mathbf{h}_{j}$ is an external (generally space and time-dependent) magnetic field. The quantum spin operators $\mathbf{S}_{i}$ have $SU(2)$ commutation relations ($a,b,c=1,2,3$)
\be
	\left[S_{j}^{a},S_{k}^{b}\right]=i\delta_{jk}\epsilon^{abc}S_{j}^{c}\,.
 \la{eq:SScom}
\ee
Commuting the spin operator $S_{j}^{a}$ with the Hamiltonian Eq.~(\ref{eq:Hmag}) one obtains Heisenberg equation of motion for the spin operator
\be
	\partial_{t}\mathbf{S}_{j} = i[H,\mathbf{S}_{j}]
	= -\sum_{k} J_{jk} \mathbf{S}_{k}\times \mathbf{S}_{j}+ \mathbf{h}_{j}\times \mathbf{S}_{j}\,.
 \la{eq:spin-eqm}
\ee

\subsection{Path integral for quantum magnets}

\subsubsection{Path integral for the magnet on a lattice}

The classical action for the magnet (\ref{eq:Hmag},\ref{eq:SScom}) can be written as 
\be
	S = -4\pi i S \sum_{j} W_{0}[\mathbf{n}_{j}] +\int d\tau\, H \,,
 \la{eq:mag-lat}
\ee
where we introduced classical unit vectors $\mathbf{n}_{i}$ and summed the terms (\ref{eq:wz0}) for each spin. The classical Hamiltonian used in (\ref{eq:mag-lat}) is obtained from (\ref{eq:Hmag}) substituting $\mathbf{S}_{i}$ by $S\mathbf{n}_{i}$. Variation of the action (\ref{eq:mag-lat}) over $n_{j}$ with the use of Eq.~(\ref{WZvar}) produces classical equation of motion 
\be
	 - i S\,  \p_{\tau}\mathbf{n}_{j}\times \mathbf{n}_{j} +
	 S^{2}\sum_{k}J_{kj}\mathbf{n}_{k} -S\sum_{i}\mathbf{h}_{j} =0\,.
 \la{eq:mag-latvar}
\ee
Taking a cross-product with $\mathbf{n}_{j}$ gives a classical analogue of (\ref{eq:spin-eqm})
\be
	  -i\p_{\tau} \mathbf{n}_{j} =
	 S\sum_{k}J_{kj}\mathbf{n}_{k}\times\mathbf{n}_{j} -\sum_{i}\mathbf{h}_{j}\times \mathbf{n}_{j} \,.
 \la{eq:spin-eqm-class}
\ee
Remember that $-i\p_{\tau}=\p_{t}$.

The path integral over trajectories of unit vectors $\mathbf{n}_{j}(\tau)$ with the amplitude  $e^{-S}$ corresponding  to the classical action (\ref{eq:mag-lat}) gives the quantization corresponding to (\ref{eq:Hmag},\ref{eq:SScom}). Our goal now is to find a continuum quantum field theory description of this lattice magnet. 

Here important remark is in order. A given lattice theory does not necessarily have a reasonable continuum description. One needs a special reason for continuum approximation to be applicable. Such reasons could be the vicinity to a second order phase transition where correlation length becomes much bigger than the lattice spacing or some other reasons for scale separation. In the following we will try to first derive a continuum limit for the theory and then check the self-consistency of the continuum approximation. Another important point is that the way to take a continuum limit depends crucially on the state of the system. In the following we first consider the ferromagnetic state and then go to the collinear antiferromagnetic state.

\subsubsection{Continuum limit for Quantum Ferromagnet}

Let us assume for simplicity that $J_{jk}=J<0$ for nearest neighbor sites $j,k$ of a spin chain in 1d, square lattice in 2d and cubic lattice in 3d and magnetic field is constant $\mathbf{h}_{j}=\mathbf{h}$. The classical Hamiltonian is then
\be
	H = -|J| S^{2} \sum_{<kj>}\mathbf{n}_{k}\cdot \mathbf{n}_{j} - S\sum_{j}\mathbf{h}\cdot\mathbf{n}_{j} \,.
 \la{eq:fm-ham}
\ee
We assume that there is a short range ferromagnetic order, i.e. nearest neighbor spins are almost perfectly aligned. We replace spins $\mathbf{n}_{j}$ at lattice sites by a continuous field $\mathbf{n}(x)$ and proceed as follows. Up to a constant, $\mathbf{n}_{j+e_{x}}\cdot \mathbf{n}_{j} \to -\frac{1}{2}(\mathbf{n}_{j+e_{x}}-\mathbf{n}_{j})^{2} \to -\frac{1}{2} a^{2} (\p_{x} \mathbf{n})^{2}$ etc. Here $a$ is the lattice constant. Replacing the summation over $j$ by the integration over space, we obtain the continuum limit of the Hamiltonian (\ref{eq:fm-ham})
\be
	H = -\frac{1}{2} |J| S^{2}a^{2} \int \frac{d^{d}x}{a^{d}}(\p_{\mu}\mathbf{n})^{2} 
	- S\int \frac{d^{d}x}{a^{d}}\mathbf{h}\cdot\mathbf{n} \,.
 \la{eq:fm-ham-cont}
\ee
Similarly, we have for the full action (\ref{eq:mag-lat})
\bea
	S[\mathbf{n}] &=& -4\pi i S \int \frac{d^{d}x}{a^{d}} W_{0}[\mathbf{n}(x,\tau)]
 \la{eq:fm-action} \\
	&-&  \frac{1}{2} |J| S^{2}a^{2} \int d\tau\frac{d^{d}x}{a^{d}}(\p_{\mu}\mathbf{n})^{2} 
	- S\int d\tau \frac{d^{d}x}{a^{d}}\mathbf{h}\cdot\mathbf{n}\,.
 \nonumber
\eea
Variation of this action with respect to the continuous unit vector field $\mathbf{n}(x,\tau)$ produces the well-known classical Landau-Lifshitz equation for magnetization (to go to real time one should replace $i\p_{\tau}\to -\p_{t}$)
\be
	i\p_{\tau}\mathbf{n} = |J| S a^{2}\, (\mathbf{n}\times \Delta \mathbf{n}) - \mathbf{h}\times \mathbf{n} \,.
 \la{eq:LLequation}
\ee

Let us remark here that $W_{0}[\mathbf{n}_{j}]$ is a topological term for an individual spin $\mathbf{n}_{j}$ on the site $j$ of the lattice. However, due to the integration over space, the first term of the continuum action Eq.~(\ref{eq:fm-action}) depends on the spatial metric (e.g., distortions of the lattice will change it). Therefore, this term cannot be considered topological. Nevertheless, it is linear in time derivative and therefore time-reparameterization invariant. Therefore, it remains imaginary after  Wick's rotation and results in a very essential interference even in imaginary time path integral.

\subsubsection{Bloch's law: the dispersion of spin waves in ferromagnet}

As an application of the continuum theory for magnetization in ferromagnets, let us derive the dispersion of spin waves starting from (\ref{eq:LLequation}). We assume that magnetic field is constant and uniform $\mathbf{h}=(0,0,h)$, and that there is a long range ferromagnetic order with spins oriented in the same direction.  We also assume that spin fluctuations are small and write $\mathbf{n} = (u_{1},u_{2},1)$, where $u_{1,2}$ are components of $\mathbf{n}$ in $x,y$ directions in spin space that are assumed to be small so that $\mathbf{n}^{2}=1+u_{1}^{2}+u_{2}^{2}\approx 1$ up to quadratic terms in $u$. Substituting all this in (\ref{eq:LLequation}) we obtain
\be
	-i\omega u  = -i |J| S a^{2}\mathbf{k}^{2}u  - i h u \,. 
\ee
Here we introduced complex notation $u=u_{1}+iu_{2}$ and made Fourier transform $\Delta \to -\mathbf{k}^{2}$ and $i\p_{\tau}\to -\p_{t} \to i\omega$. We immediately obtain the dispersion of spin waves 
\be
	\omega =  |J| S a^{2}\mathbf{k}^{2} +h\,,
\ee
the result known as the Bloch's law. In the absence of external magnetic field the dispersion of spin waves is quadratic in wave vector $k$.
 
To conclude our brief discussion of the ferromagnetic case we have to recall that the continuum theory was derived under the condition that fluctuations of $\mathbf{n}$ are small compared to $1$ or $|u|\ll 1$. Given a temperature and other parameters of the theory one should calculate the average value of those fluctuations. The condition $\langle |u|^{2}\rangle \ll 1$ is then the necessary condition for the self-consistency of the continuum approximation.

\subsection{Continuum path integral for Quantum Antiferromagnet}
\la{sec:cpiQAFM}

Let us consider a more subtle case of quantum antiferromagnet. We again start with the Hamiltonian  (\ref{eq:fm-ham}). However, we assume now that $J>0$ and write 
\be
	H = J S^{2} \sum_{<kj>}\mathbf{n}_{k}\cdot \mathbf{n}_{j} - S\sum_{j}\mathbf{h}\cdot\mathbf{n}_{j} \,.
 \la{eq:afm-ham}
\ee
Although (\ref{eq:afm-ham}) looks very similar to (\ref{eq:fm-ham}) the unit vectors $\mathbf{n}_{j}$ tend to be antiparallel on nearest sites (we again assume square lattice here so that the antiferromagnetic order is not frustrated). One cannot use the continuous field $\mathbf{n}(x)$ instead of lattice vectors $\mathbf{n}_{j}$. Taking continuum limit is more involved and can be achieved through the following substitution
\be
	\mathbf{n}_{j} = (-1)^{j} \mathbf{m}(x) + a \mathbf{l}(x)\,.
 \la{eq:ml}
\ee
Here we assume that both fields $\mathbf{m}(x)$ and $\mathbf{l}(x)$ are good continuous (smooth) fields. \footnote{We emphasize that in the following we take a particular continuum limit which assumes short range ordered antiferromagnetic state. It is believed to be appropriate for large $S$ Heisenber antiferromagnets. However, it is not appropriate, e.g. for spin chains at so-called Bethe Ansatz integrable points.} The former represents the smooth staggered magnetization while the latter is a ferromagnetic component. It is expected that the ferromagnetic component is small and the corresponding rescaling by the lattice constant $a$ is made. As $\mathbf{n}_{j}^{2}=1$ we have
\be
	\mathbf{n}_{j}^{2}=\mathbf{m}^{2} + 2(-1)^{j}a (\mathbf{m}\cdot\mathbf{l}) +a^{2}\mathbf{l}^{2} =1\,.
\ee
We solve this condition to the order of $a^{2}$ by two conditions
\be
	\mathbf{m}^{2}=1\,, \qquad\quad \mathbf{m}\cdot\mathbf{l} =0\,.
 \la{eq:ml-constraint}
\ee
Using (\ref{eq:ml}) we have up to constants
\bea
	\mathbf{n}_{j+e_{x}}\cdot \mathbf{n}_{j} 
	&\to &  \frac{1}{2}a^{2}\left[(\p_{x}\mathbf{m})^{2} 
	+4\mathbf{l}^{2}+4(-1)^{j}\p_{x}\mathbf{m}\cdot\mathbf{l}\right] \,,
 \nonumber \\
 	\mathbf{h}\cdot\mathbf{n}_{j}
	&\to & (-1)^{j}\mathbf{h}\cdot\mathbf{m} + a\mathbf{h}\cdot\mathbf{l} \,.
 \nonumber
\eea
Substituting these expressions into (\ref{eq:afm-ham}) we obtain
\bea
	H &=& J S^{2} \sum_{j}\frac{1}{2}a^{2}\left[(\p_{\mu}\mathbf{m})^{2} 
	+4d\mathbf{l}^{2}+4(-1)^{j}\p_{\mu}\mathbf{m}\cdot\mathbf{l}\right] 
	- S\sum_{j}((-1)^{j}\mathbf{h}\cdot\mathbf{m} + a\mathbf{h}\cdot\mathbf{l}) \,.
 \nonumber \\
 	&\to & J S^{2} a^{2}\frac{1}{2} \int \frac{d^{d}x}{a^{d}} 
	\left[(\p_{\mu}\mathbf{m})^{2} +4d\mathbf{l}^{2}\right] 
	-Sa \int \frac{d^{d}x}{a^{d}} \mathbf{h}\cdot\mathbf{l} \,.
\eea
In the last step, we dropped all oscillating terms and replaced summation by integration over space.

The next step is to do a similar procedure with the term in the action coming from the summation of topological terms. We proceed as follows
\bea
	\sum_{j} W_{0}[\mathbf{n}_{j}] 
	 &=&  \sum_{j} W_{0}[(-1)^{j} \mathbf{m}(x) + a \mathbf{l}(x)]
 \nonumber \\
	 &\approx & \sum_{j} (-1)^{j} W_{0}[\mathbf{m}(x) + (-1)^{j}a \mathbf{l}(x)]
 \nonumber \\
	&\approx &  \sum_{j} (-1)^{j} W_{0}[\mathbf{m}(x)] 
	+ \int \frac{d^{d}x}{a^{d}}\int d\tau\,a \mathbf{l}(x) \frac{\delta W_{0}[\mathbf{m}]}{\delta\mathbf{m}}
 \nonumber
\eea
We now use the variation formula (\ref{WZvar}) and its consequence
$$
	W_{0}[\mathbf{m}(x+e_{x})]-W_{0}[\mathbf{m}(x)]
	\approx \frac{1}{4\pi}\int d\tau\, (a\p_{x}\mathbf{m})\cdot(\mathbf{m}\times \p_{\tau}\mathbf{m}) 
$$
and obtain
\bea
	 \sum_{j} W_{0}[\mathbf{n}_{j}] 
	&\approx & \frac{1}{4\pi} \int \frac{d^{d}x}{a^{d}}
	\int d\tau\, a d\; \mathbf{l}(x)\cdot (\mathbf{m}\times \p_{\tau}\mathbf{m})
 \la{eq:topcontafm} \\
	&+&   \sum_{j_{y},j_{z}} (-1)^{j_{y}+j_{z}} \frac{1}{8\pi} \int \frac{dx}{a}
	\int d\tau\, (a\p_{x}\mathbf{m})\cdot(\mathbf{m}\times \p_{\tau}\mathbf{m}) \,.
 \nonumber
\eea
The first term of (\ref{eq:topcontafm}) is written for any spatial dimension $d$. In the second term, we assumed the three-dimensional case. Notice that while the sign alternation was taken into account in $x$ direction there is still a sum to be taken with the factor $(-1)^{j_{y}+j_{z}}$ in other two directions. That summation will suppress this term and, therefore, it is relevant only in one spatial dimension. We summarize for the topological contribution
\bea
	 -4\pi i S \sum_{j} W_{0}[\mathbf{n}_{j}] 
	&\approx & -i S a^{1-d}d 
	\int d\tau\,d^{d}x\;  \mathbf{l}(x)\cdot (\mathbf{m}\times \p_{\tau}\mathbf{m})
 \nonumber \\
	&-&  i \frac{S}{2} \delta_{d,1}
	\int d\tau\,dx\, \p_{x}\mathbf{m}\cdot(\mathbf{m}\times \p_{\tau}\mathbf{m}) \,.
 \la{eq:topcontafm1}
\eea

Collecting all terms together to get the continuum limit of the action (\ref{eq:mag-lat}) we obtain
\bea
	S[\mathbf{m},\mathbf{l}] &=& i \frac{S}{2} \delta_{d,1}
	\int d\tau\,dx\, \p_{x}\mathbf{m}\cdot(\mathbf{m}\times \p_{\tau}\mathbf{m})
	-i S a^{1-d}d 
	\int d\tau\,d^{d}x\;  \mathbf{l}(x)\cdot (\mathbf{m}\times \p_{\tau}\mathbf{m})
 \nonumber \\
	&+&
	J S^{2} a^{2-d}\frac{1}{2} \int d\tau\,d^{d}x\,
	\left[(\p_{\mu}\mathbf{m})^{2} +4d\mathbf{l}^{2}\right] 
	-Sa^{1-d} \int d\tau\,d^{d}x\, \mathbf{h}\cdot\mathbf{l}
 \nonumber \\
 	&=& i \frac{S}{2} \delta_{d,1}
	\int d\tau\,dx\, \p_{x}\mathbf{m}\cdot(\mathbf{m}\times \p_{\tau}\mathbf{m})
	+J S^{2} a^{2-d}\frac{1}{2} \int d\tau\,d^{d}x\,
	(\p_{\mu}\mathbf{m})^{2}
 \nonumber \\
 	&+&  \int d\tau\,\frac{d^{d}x}{a^d}\,\left( 2J S^{2} a^{2} d {\mathbf{l}}^{2}
	-Sa^{1}  \mathbf{l} \cdot \Big[\mathbf{h} 
	+i  d\, (\mathbf{m}\times \p_{\tau}\mathbf{m})\Big]\right)\,.
 \la{eq:Sml}
\eea
The obtained expression is the continuum limit of (\ref{eq:mag-lat}) derived in the antiferromagnetic regime with the assumption of small fluctuations around the short range collinear antiferromagnetic order. The field $\mathbf{l}$ describing the magnetization of the magnet enters the action in a very simple way and can be ``integrated out''. For details of derivation see the Appendix~\ref{app:l-derivation}. Here we present the results dropping the external magnetic field for simplicity. In two and higher spacial dimensions $d>1$, we have
\bea
	S[\mathbf{m}] &=& \frac{1}{2g}  \int d\tau\,\frac{d^{d}x}{a^{d-1}}\,
	\left[\frac{1}{v_{s}}(\p_{\tau}\mathbf{m})^{2}+v_{s}(\p_{\mu}\mathbf{m})^{2}\right]\,,
 \la{eq:NLSMafm}
\eea
where 
\be
	v_{s} = \frac{2JSa}{\sqrt{d}}\,, \qquad g=\frac{2}{S\sqrt{d}}\,.
\ee

The one-dimensional case is special and has an additional topological term in the action
\bea
	S[\mathbf{m}] &=& \frac{1}{2g}  \int d\tau\,dx\,
	\left[\frac{1}{v_{s}}(\p_{\tau}\mathbf{m})^{2}+v_{s}(\p_{x}\mathbf{m})^{2}\right]
 \nonumber \\
	&+& i\theta \int d\tau\,dx\, \frac{1}{4\pi} \mathbf{m}\cdot(\p_{\tau}\mathbf{m}\times \p_{x}\mathbf{m})\,, 
 \la{eq:NLSMafmtheta}
\eea
where 
\be
	v_{s} = 2SaJ\,, \qquad g=\frac{2}{S}\,, \qquad \theta =2\pi S\,.
 \la{eq:nlsm-par}
\ee
The model (\ref{eq:NLSMafmtheta}) is known as $O(3)$ nonlinear sigma model with topological theta-term. It is the low energy, long distance description of the antiferromagnetic Heisenberg spin chain with large spin $S\gg 1$ with the correspondence between parameters of the model and the parameters of the spin chain given by (\ref{eq:nlsm-par}). Let us start with the discussion of the nonlinear sigma model without topological term.

\subsection{RG for O(3) NLSM}

The model, Eq.~(\ref{eq:NLSMafmtheta}) without topological term ($\theta=0$) can be re-written as
\bea
	S[\mathbf{m}] &=& \frac{1}{2g}  \int d^2x\, (\p_{\mu}\mathbf{m})^{2}\,,
 \la{eq:O3NLSM}
\eea
where $\mu=\tau,x$ and we re-defined $\tau\to \tau/v_{s}$. The action (\ref{eq:O3NLSM}) with the constraint $\mathbf{m}^{2}=1$ is known as $O(3)$ nonlinear sigma model (NLSM). It is relativistically invariant with spin wave velocity $v_{s}$ playing the role of the speed of light. This relativistic invariance is emergent and we should remember that the next order gradient corrections to the model and various perturbations are generally not relativistically invariant. 

At small values of the coupling constant $g$ corresponding to large values of $S$ one can treat (\ref{eq:O3NLSM}) perturbatively and ask how the coupling constant renormalizes when one goes to longer distances. It turns out \cite{polyakov1975interaction} that $g$ increases with the scale. The increase of $g$ signals the tendency of the $\mathbf{m}$-field to disorder. More precisely, the effective coupling of (\ref{eq:O3NLSM}) at the length $L$ satisfies renormalization group (RG) equation 
\be
	\frac{dg}{d\log L} = \frac{1}{2\pi}g^2 +O(g^3)\,, 
 \la{eq:O3RG}
\ee
and gives
\be
	g(L) = \frac{g_0}{1-\frac{g_0}{2\pi}\log (L/a)} \,,
\ee
where $g_0=g(a)$ is the coupling constant at UV (lattice) scale $a$. At the scale $L\sim \xi$ with 
\be
	\xi \sim a e^{2\pi/g_0}\,,
 \la{eq:O3xi}
\ee
the effective coupling constant $g(\xi)$ becomes of the order of unity and we cannot trust RG equation (\ref{eq:O3RG}) at this point. 

We stress here that RG analysis is not conclusive. The only conclusion we can make is that the effective length $\xi$ given by (\ref{eq:O3xi}) emerges. At this scale, the $\mathbf{m}$ field is somewhat disordered, but we cannot say anything about the nature of the phase and about the long distance behavior of $\mathbf{m}$-field correlation functions. There are essentially two scenarios. The first one is that the actual model has a gap of the order of $v_s/\xi$, the field $\mathbf{m}$ is disordered with all correlations decaying exponentially with the correlation length $\xi$ (\ref{eq:O3xi}). The second possibility is that the RG flow leads to a new fixed point and behavior of the model at scales larger than $\xi$ is governed by that fixed point (in particular, long range correlation functions might decay as power laws etc.). It turns out that this is the former scenario that is realized for 2d $O(3)$ nonlinear sigma model (\ref{eq:O3NLSM}). We know this because $O(3)$ NLSM has been solved exactly by Bethe Ansatz \cite{wiegmann1985exact} and has a gap separating the ground state from excitations.  In the next section we will argue that the second scenario might be relevant when topological term is present in NLSM.

Bethe Ansatz solution of the model (\ref{eq:O3NLSM}) is outside of the scope of these lectures. Instead, to have some understanding of how finite gap (correlation length) appears in NLSM we refer the reader to the Exercise \ref{ex:ONNLSM} ``O(N) NLSM'' below where the correlation length is obtained for the $O(N)$ NLSM in the limit of large $N$.

\subsection{O(3) NLSM with topological term}
\la{sec:O3top}

Let us now consider the O(3) NLSM with topological theta term
\bea
	S[\mathbf{m}] &=& \frac{1}{2g}  \int d^2x\, (\p_{\mu}\mathbf{m})^{2}
	+i\theta Q\,,	
 \la{eq:O3NLSMtheta}
\eea
where 
\bea
	Q &=& \int d^{2}x\, \frac{1}{4\pi} \mathbf{m}\cdot(\p_{\tau}\mathbf{m}\times \p_{x}\mathbf{m})\,.
 \la{eq:Qexp}
\eea
We assume that the boundary conditions $\mathbf{m}(x) \to \mathbf{m}_{0}=const$,  as $x\to \infty$ so that the winding number $Q$ is an integer. The parameters corresponding to the AFM spin chain are given by (\ref{eq:NLSMafmtheta},\ref{eq:nlsm-par}), i.e., $g=2/S$ and $\theta =2\pi S$. Following Haldane \cite{haldane1983nonlinear,haldane1983continuum}, we notice that the topological term in (\ref{eq:O3NLSMtheta}) contributes the complex weight to path integral given by $e^{i\theta Q}=(-1)^{2S Q}$. This weight depends crucially on the integer-valuedness of spin. If the spin $S$ is half-integer, the weight is non-trivial $(-1)^{Q}$ and results in interference of topological sectors characterized by different topological charges $Q$. On the other hand, if $S$ is integer, the weight is unity and does not affect the path integral. \footnote{This statement is not quite correct. The topological term can still be important for various boundary conditions and due to the presence of singularities.} Based on this observation Haldane conjectured that AFM spin chains with integer spin $S$ have singlet ground states separated by finite gap from all excitations similar to O(3) NLSM without topological term. On the other hand, AFM spin chains with half-integer spin have gapless excitations similar to the spin-1/2 chain. For the latter, the spectrum of excitations has been known from the exact solution by Bethe \cite{bethe1931theorie}. The Haldane's conjecture has been supported by numerical simulations and experiments. 

It is instructive to think about RG flow for the NLSM with theta term. The model has two parameters $g$ and $\theta$ and it is appealing to think about RG flow in the plane labeled by those parameters. First of all, we notice that starting with dimerized spin chain one obtains values of $\theta$ which are not necessarily multiples of $\pi$ (see Exercise \ref{ex:dimspin} ``Dimerized spin chain'' below). Therefore, it is tempting to draw the flow diagram similar to the one for integer quantum Hall effect \cite{levine1983electron,khmel1983quantization} (see figure \ref{fig:rgflow}). 
\begin{figure}
\bigskip
\begin{center}
 \includegraphics[width=12cm]{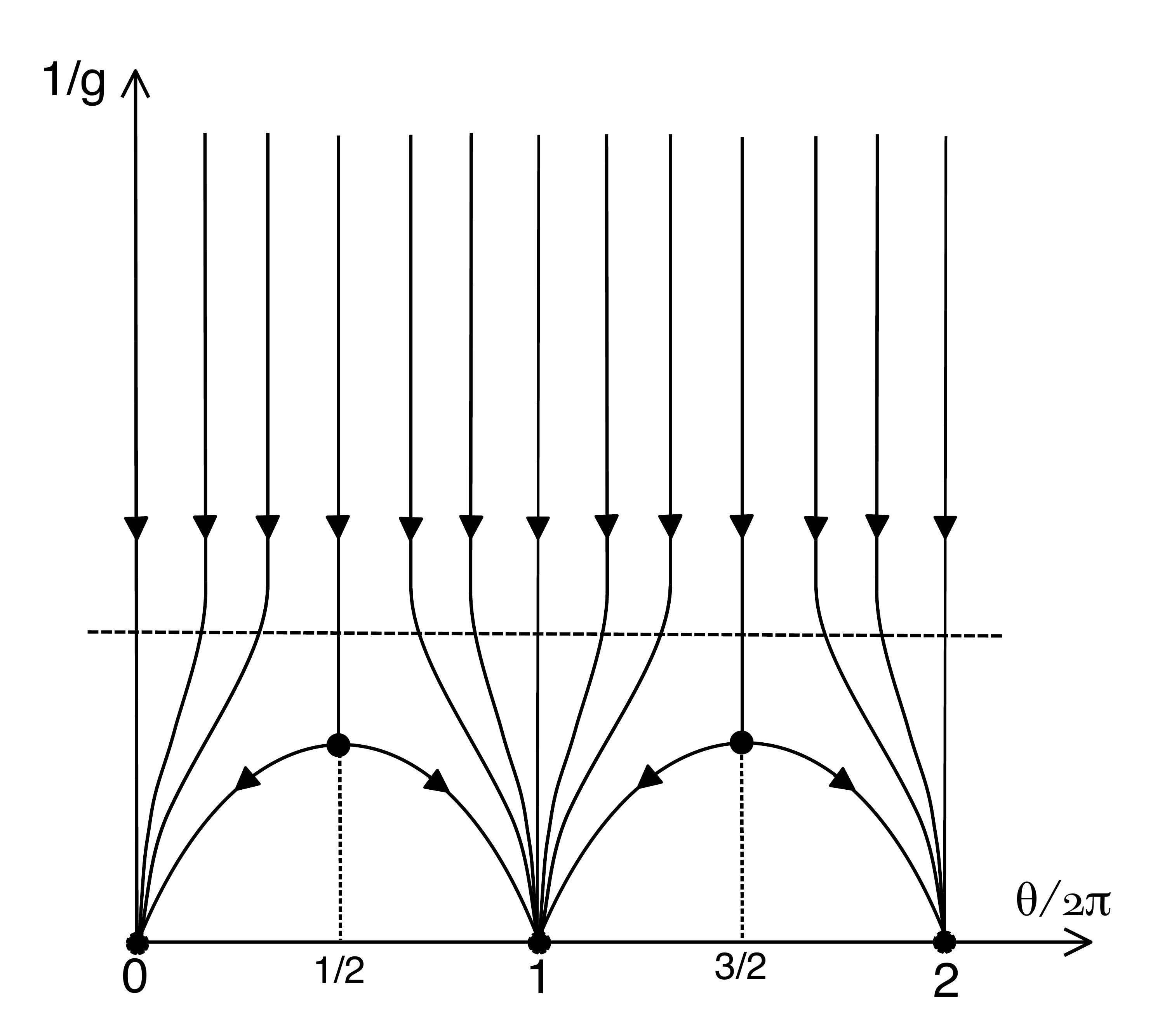}
\end{center}
\caption{The schematic RG flow for nonlinear sigma model with theta term is shown. The vertical axis is the inverse coupling constant $g$ while the horizontal axis is the topological angle $\theta$. The picture has exact symmetry with respect to $\theta\to\theta+2\pi$ and with respect to reflection relative to half integer $\theta/2\pi$. The horizontal dashed line corresponds to $g\sim 1$. Below this line the perturbative RG is not working and interpretation of the flow is more subtle.}
\la{fig:rgflow}
\end{figure}

It is not difficult to see that conventional perturbative calculation results in non-renormalizability of $\theta$. This is reflected by strictly vertical flow at small $g$ (large $1/g$) in figure \ref{fig:rgflow}. One can argue following \cite{levine1983electron} that taking into account instanton configurations (configurations with $Q\neq 0$) will result in the deviations of flow from the vertical one as shown in the figure. However, the instanton contribution is suppressed by factors of the order of $e^{-2\pi/g}$ and it is very hard to develop consistent perturbation theory taking into account exponentially small terms but neglecting terms of higher order in coupling $g$. More straightforward interpretation of figure \ref{fig:rgflow} is that it shows the flow of some observables which correspond to $1/g$ and $\theta$ at small values of $g$ and change with scale as shown in the figure (in analogy with observable conductances $\sigma_{xx}$ and $\sigma_{xy}$ of \cite{levine1983electron}). 
We will not dwell on the interpretation of figure (\ref{fig:rgflow}). However, it is believed that as shown on that figure the long distance behavior of spin chains of half-integer $\theta/(2\pi)$ is governed by new infrared fixed points while the spin chain with any other $\theta$ flows to the models with finite gap in the spectrum. In fact, it was clarified that the critical description of half-integer spin chains is given by Wess-Zumino-Witten (WZW) model (see Ref.~\cite{1987-AffleckHaldane}). In particular, the staggered magnetization correlation decays as power law at large distances.

In the limit of large $S$, we have the following picture for the correlator $\langle \mathbf{m}(x)\cdot\mathbf{m}(0)\rangle$. At distances shorter than $\xi \sim a e^{\pi S}$ (obtained form (\ref{eq:O3xi}) at $g=2/S$) there is a short range antiferromagnetic order $\langle \mathbf{m}(x)\cdot\mathbf{m}(0)\rangle \sim 1$. At larger distances $x\gg \xi$, the order is destroyed, and there is no long-range order in agreement with the Mermin-Wagner theorem. However, the way in which correlations decay at large distances depend crucially on $S$. If $S$ is integer, the decay is exponential $\langle \mathbf{m}(x)\cdot\mathbf{m}(0)\rangle \sim e^{-x/\xi}$, while at half-integer $S$, it is algebraic (power-law) with $\langle \mathbf{m}(x)\cdot\mathbf{m}(0)\rangle \sim x^{-1}$. \footnote{At non-zero temperature there is another length scale $\xi_T \sim a JS^2/T$, and power-law decay of correlations will be eventually replaced by exponential decay at large distances due to thermal fluctuations.} This power-law signals the emergent scale invariance (in fact, conformal invariance) and is captured by the infrared fixed point of effective WZW model. Interestingly, for half-integer spins the symmetry of the spin chain is enlarged at the critical point. In addition to the fluctuations of the staggered magnetization $\mathbf{m}(x)$, fluctuations of dimerization become soft as well. They combine into the effective $SU(2)$ ``order parameter'' which is an $SU(2)$ degree of freedom of WZW model. We refer the reader to \cite{1987-AffleckHaldane} for more detailed analysis of critical behavior of spin chains.

\subsection{Boundary states for spin 1 chains with Haldane's gap}
\la{sec:bound-states-haldane}

We argued at the beginning of Sec.~\ref{sec:O3top} that for integer $S$ the topological term is ``ineffective'' and the model (\ref{eq:O3NLSMtheta}) behaves similarly to the O(3) NLSM without topological term (\ref{eq:O3NLSM}). This is not quite correct. The argument is essentially based on the integer-valuedness of the topological charge $Q$. However, it is necessarily integer only for smooth $\mathbf{m}$-field configurations for compact boundary conditions (e.g., $\mathbf{m}\to \mathbf{m}_0$ at $x,t\to \infty$). Let us consider an example when boundary conditions are different.

We assume that the spin chain is long but has a finite length $L$ with free boundary conditions (boundary spins at $x=0, L$ can take arbitrary values). Then the topological term can be written as
\bea
	S_{top}[\mathbf{m}] &=&  i\theta \int d\tau\, \int_0^L dx\, \frac{1}{4\pi} \mathbf{m}\cdot(\p_{\tau}\mathbf{m}\times \p_{x}\mathbf{m})
 \nonumber \\
 	&=& i\theta \left[\frac{\Omega_L}{4\pi}+Q-\frac{\Omega_0}{4\pi}\right]\,. 
 \la{eq:topboundary}
\eea
Here we decomposed the winding number (\ref{eq:Qexp}) into integer part $Q$ and the difference of solid angles $\Omega_0,\Omega_L$ subtended by boundary vectors $\mathbf{m}(0)$ and $\mathbf{m}(L)$, respectively. Assuming $\theta =2\pi$, we drop the integer part and end up with the contribution
\bea
	S_{top} = i \left[\frac{\Omega_L}{2}-\frac{\Omega_0}{2}\right]\,,
 \la{eq:sbound}
\eea 
which is written in terms of boundary vectors only. Comparing with (\ref{seff1loc}) we recognize (\ref{eq:sbound}) as the action of two spin-1/2 located at the ends of the spin chain. As we expect that the bulk degrees of freedom have a gap in the spectrum ($S=1$ is an integer) we conclude that the $S=1$ AFM spin chain in the gapful phase (Haldane's phase) should have gapless boundary spin 1/2 excitations. 

We would like to stress again how unusual is the conclusion we have just made. It is well known that ``adding'' finite number of spin 1s one can get only superposition of integer valued spins. For example, two spin-1 particles can only have sectors with total spins $0,1$, and $2$. However, we managed adding large number of spin 1s to get two boundary spin 1/2 as low lying excitations of the finite spin chain with free boundary conditions! \footnote{Actually, for the finite length of the spin chain those two boundary spin 1/2 are effectively interacting with the strength of interaction $\sim e^{-L/\xi}$. Only in the limit of an infinite length we obtain truly non-interacting spin 1/2 degrees of freedom. } Most amazingly, these boundary spin 1/2 have been observed in experiment. \cite{hagiwara1990observation,kenzelmann2003structure}
\begin{figure}
\bigskip
\begin{center}
 \includegraphics[width=8cm]{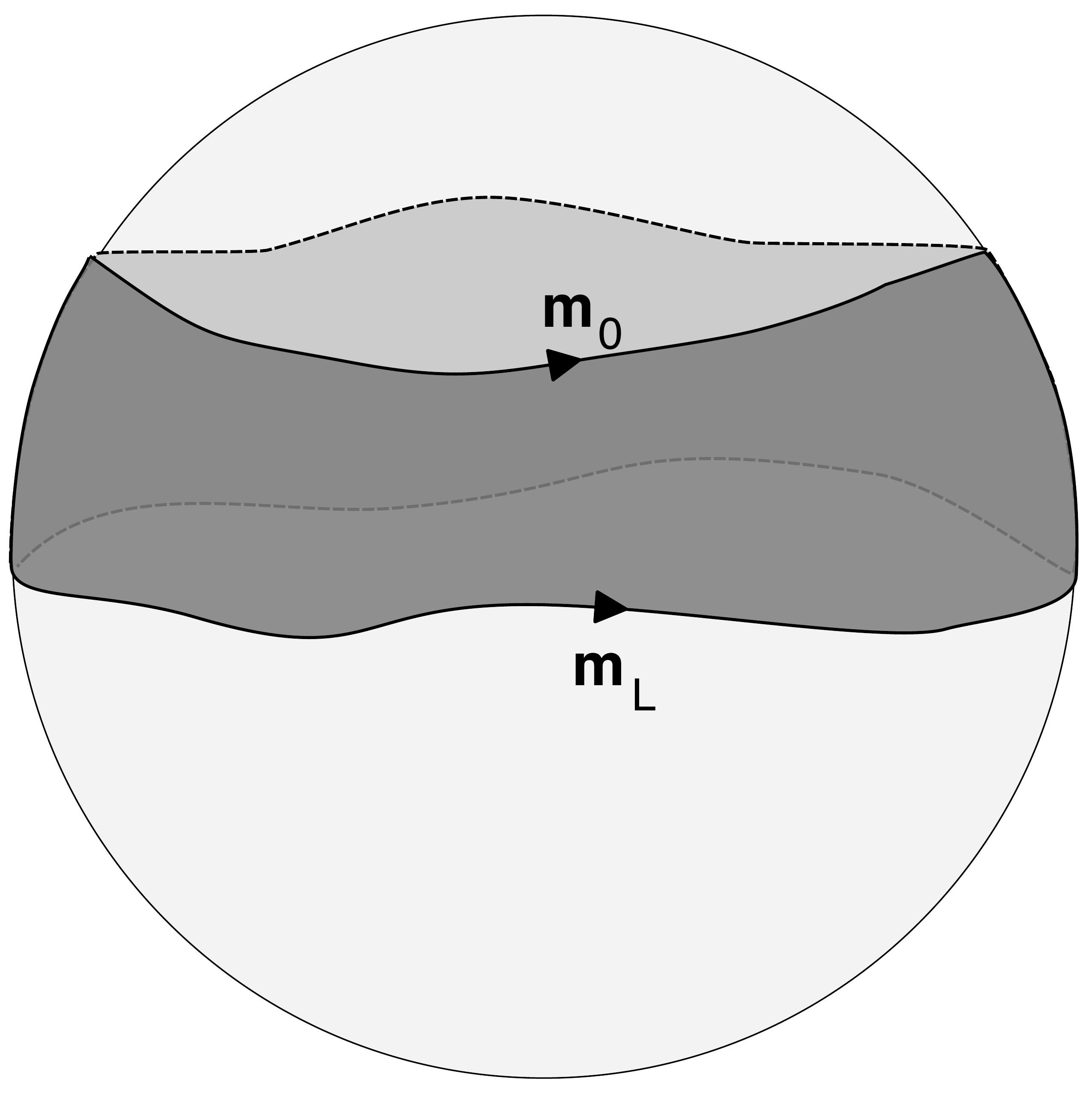}
\end{center}
\caption{Time evolution of boundary magnetizations (unit vectors) $\mathbf{m}_{0}(\tau)$ and $\mathbf{m}_{L}(\tau)$ is shown. For periodic time these vectors sweep solid angles $\Omega_0$ and $\Omega_L$, respectively. The shaded area is given by (\ref{eq:topboundary}) which is up to integer $Q$ is given by the difference of solid angles $\Omega_L$ and $\Omega_0$ as in (\ref{eq:sbound}).}
\la{fig:boundaryspins}
\end{figure}

This example illustrates very interesting connection between $\theta$ and WZW topological terms. The gapful model with $\theta$ terms might produce massless boundary theory of one dimension lower described with the use of WZ terms.

\subsection{AKLT model}

To understand better how the bulk gap and boundary spin 1/2 states are formed in the Haldane's phase of $S=1$ antiferromagnetic spin chain let us consider the deformation of the Heisenberg Hamiltonian known as AKLT model \cite{affleck1987rigorous,affleck1988valence}
\bea
	H = J \sum_j \left(\mathbf{S}_j\cdot\mathbf{S}_{j+1} + \frac{1}{3}(\mathbf{S}_j\cdot\mathbf{S}_{j+1})^2\right) \,.
 \la{eq:AKLT}
\eea 
One should think about this model as fine tuned model with the coefficient $J'$ in front of the biquadratic term being exactly 1/3 of the coefficient in front of Heisenberg exchange term. Changing $J'$ from $0$ to $J/3$ interpolates between Heisenberg spin chain and AKLT model (\ref{eq:AKLT}). It can be shown that the ground state of (\ref{eq:AKLT}) is separated by the finite gap from excitations. If this gap is not closed in the process of changing $J'$ from $J/3$ to zero then the Heisenberg spin chain has a gap as well and we say that AKLT and Heisenberg model are adiabatically connected and are in the same phase. 

Let us now see why AKLT model is much easier to analyze than the Heisenberg model for $S=1$. It turns out that the Hamiltonian (\ref{eq:AKLT}) up to constant can be written as
\bea
	H = 2J \sum_j P_2(\mathbf{S}_j+\mathbf{S}_{j+1}) \,,
 \la{eq:AKLT2}
\eea
where 
\be
	P_2(\mathbf{S}_1+\mathbf{S}_2) 
	= \frac{1}{2}\mathbf{S}_j\cdot\mathbf{S}_{j+1} 
	+ \frac{1}{6}(\mathbf{S}_j\cdot\mathbf{S}_{j+1})^2+\frac{1}{3}
 \la{eq:proj2}
\ee
is a projector on the state with the total spin $S=2$ (see the Exercise \ref{ex:projector} ``Projector to $S=2$'').
Then schematically (see \cite{affleck1987rigorous,affleck1988valence} for details) one can think about the ground and excited states of (\ref{eq:AKLT2}) in the following way. We split every spin $S_j$ into two spins 1/2 (let us call them $A$ and $B$. Taking one of those two spin 1/2 from the site $j$ and one of spin 1/2 from the site $j+1$ we form a singlet state and make a product of all these singlets. Then we project the result back to the total spin 1 on each site. The obtained state
$${\cal P}\prod_j \Big(|+\rangle_{j,A}|-\rangle_{j+1,B}-|-\rangle_{j,A}|+\rangle_{j+1,B}\Big)$$ 
has a property that the sum of two neighbor spins is never 2 and is, therefore, the ground state of (\ref{eq:AKLT2}) and of (\ref{eq:AKLT}). \footnote{Of course, these speculations are not rigorous at all. For rigorous treatment see \cite{affleck1987rigorous,affleck1988valence} or modern treatment using the technique of matrix product states (MPS).}  It is also physically clear that to excite this system one should break one of the singlets and pay finite energy. Schematically we illustrate the construction in figure \ref{fig:AKLT}. In particular, one can see that the boundary spin 1/2 in this case are just uncoupled ``halves'' of the end spins of the chain. In some sense AKLT provides a ``chemical'' model with singlets being covalent bonds and spin 1/2-s being physical electrons. We refer to AKLT model as to a strong coupling limit of the Haldane's phase of a spin-1 chain.
\begin{figure}
\bigskip
\begin{center}
 \includegraphics[width=10cm]{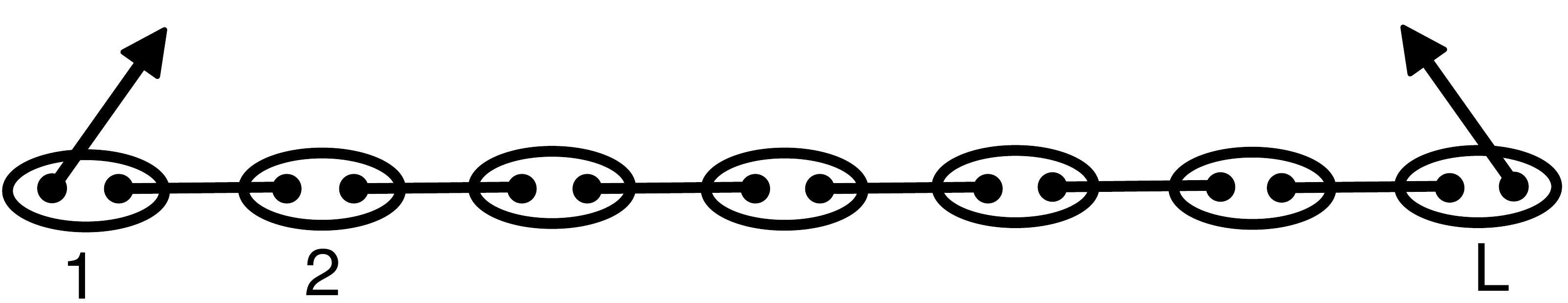}
\end{center}
\caption{The schematic picture of the ground state of AKLT model. Ovals represent the sites of the spin chain. Spin 1 at each site is split in two spin 1/2s. The horizontal segments represent singlet states of spin 1/2-s from neighbor sites. Notice that one of spin 1/2-s at the site 1 and at the site $L$ are not parts of singlets. They are the boundary spin 1/2 states of the AKLT spin 1 chain.}
\la{fig:AKLT}
\end{figure}

\subsection{Exercises}


{\small


\prn{ex:dimspin}
\subsubsection*{Exercise \ref{ex:dimspin}: Dimerized spin chain}

Start with the spin-chain Hamiltonian 
$$H= \sum_k J_k\, \mathbf{S}_k\cdot\mathbf{S}_{k+1}$$ 
with $J_k=J>0$ if $k$ even and $J_k=J'>0$ if $k$ odd. Repeat the derivation of the section \ref{sec:cpiQAFM} and show that the continuum limit is still given by $O(3)$ NLSM with theta term (\ref{eq:NLSMafmtheta}) but instead of (\ref{eq:nlsm-par}) the parameters of the model are given by
\be
	v_{s} = 2Sa\sqrt{JJ'}\,, \qquad g=\frac{1}{S}\left(\sqrt{\frac{J}{J'}}+\sqrt{\frac{J'}{J}}\right)\,, 
	\qquad \theta =2\pi S\left[1-\frac{J-J'}{J+J'}\right]\,.
 \la{eq:nlsm-parDim}
\ee
Notice that the model is self-dual with respect to the duality transformation $J\leftrightarrow J'$, $\theta\to 4\pi S-\theta$ corresponding to the reflection symmetry of the 1d lattice and the Hamiltonian.
	 
\prn{ex:ONNLSM}
\subsubsection*{Exercise \ref{ex:ONNLSM}: $O(N)$ NLSM}

Let us consider the model given by the action (\ref{eq:O3NLSM}), where the unit vector field $\mathbf{m}^2=1$ has $N$ components $\mathbf{m} = (m_1,m_2,\ldots,m_N)$. This model is known as $O(N)$ nonlinear sigma model. We replace (\ref{eq:O3NLSM}) by 
\bea
	S[\mathbf{m},\lambda] &=&   \frac{1}{2g_0} \int d^2x\, \left[(\p_{\mu}\mathbf{m})^{2}
	+i\lambda (\mathbf{m}^2-1)\right]\,.
 \la{eq:ONNLSM}
\eea
We introduce here the field $\lambda(x,t)$ so that functional integration over it gives the local constraint $\mathbf{m}^2=1$. Assuming that the functional integral over $\lambda(x,t)$ is performed exactly one should not worry about the constraint in (\ref{eq:ONNLSM}). Let us now make an approximation assuming that the functional integral corresponding to (\ref{eq:ONNLSM}) is dominated by the saddle point $i\lambda = M^2=const$. Then the path integral over $\mathbf{m}$ is Gaussian.\footnote{At this step we essentially replace the local constraint $\mathbf{m}^2=1$ for the global one $\int d^2x\, \mathbf{m}^2 = const$ and hope that fluctuations of local magnitude of $\mathbf{m}$ are not important.} 
 
Compute this Gaussian integral and find $\langle \mathbf{m}^2\rangle$ for a given $M$. Assume that all divergent integrals can be cut off by the lattice scale $a$. Write down the consistency equation $\langle \mathbf{m}^2\rangle=1$. This is the so-called \textit{gap equation} which determined the saddle point value of $M$. Show that it is given by
\bea
	M \sim \frac{1}{a} \exp\Big(-\frac{4\pi}{Ng_0}\Big)\,.
 \la{eq:Mvalue}
\eea 
Consider now fluctuations of $\lambda$ around the saddle point, and show that these fluctuations are suppressed by the parameter $1/N$, i.e. one needs large $N$ limit to make saddle point approximation self-consistent. 

As a result of this exercise we obtained that the correlation functions of $\mathbf{m}$-field behave as the ones for the field with the mass $M$ given by (\ref{eq:Mvalue}), i.e., $\langle \mathbf{m}(x)\cdot\mathbf{m}(0)\rangle \sim \exp(-Mx)$.

For the answer to this exercise see \cite{PolyakovBook-1987}.

\prn{ex:bound-spins}
\subsubsection*{Exercise \ref{ex:bound-spins}: Boundary spin 1/2 states of a Haldane's chain}

Consider the ``action'' of a two-dimensional O(3) non-linear sigma model with topological term
\begin{eqnarray}
	S &=& S_{NLSM} +S_{\theta} \,,
 \label{actsch}\\
	S_{NLSM} &=& \frac{1}{2g} \int d^{2}x\, (\partial_{\mu}\mathbf{n})^{2} \,,
 \\
 	S_{\theta} &=& i\theta Q \,,
 \la{eq:Stheta}\\
 	    Q &=& \int d^{2}x\, \frac{1}{8\pi}
    \epsilon^{\mu\nu}\mathbf{n}[\partial_{\mu}\mathbf{n}\times\partial_{\nu}\mathbf{n}] \,.
 \la{eq:Q}
\end{eqnarray}
This action can be derived as a continuum limit of the Heisenberg spin chain with large spins on sites. In the latter case, $g=2/S$ and $\theta =2\pi S$. In the case of integer $S$, the spin chain is massive, and there are no bulk excitations at low energies (smaller than the gap). 

Let us assume that the action (\ref{actsch}) is defined on the open chain of the length $L$. Show that the topological theta-term (\ref{eq:Stheta},\ref{eq:Q}) formally defined on the open chain reduces to two WZ (0+1-dimensional) terms at the boundary of spacetime, i.e. at the ends of the spin chain. This means that we expect two quantum spins living at the ends of the spin chain. Show that the coefficient in front corresponds to the value of those spins $S/2$. In particular, it means that the boundary states of $S=1$ spin chain correspond to spin-$1/2$!

\textit{Remark}: Neglecting the NLSM part of the action is possible in this exercise only because of the gap in the bulk at the integer value of spin.  

For the answer see Sec.~\ref{sec:bound-states-haldane}.

\prn{ex:projector}
\subsubsection*{Exercise \ref{ex:projector}: Projector to $S=2$}

Show that the operator (\ref{eq:proj2}) projects any state of two spins $S=1$ onto the state with total spin $S=2$.

Hint: use the identity $(\mathbf{S}_1\cdot\mathbf{S}_2) = \frac{1}{2}\left[(\mathbf{S}_1+\mathbf{S_2})^2-4\right]$ for two spin-1 operators $\mathbf{S}_1^2=\mathbf{S}_2^2=S(S+1)=2$.

}

\section{Conclusion}

In these lectures, we considered a few examples of topological terms that appear in effective actions used in condensed matter theory. We discussed some of the properties of those terms and their physical consequences. We also showed how these terms can be generated by fermionic degrees of freedom as phases of fermionic determinants. Of course, these lectures can serve only as an introduction to a quickly developing field of topological phases of matter. Because of limited time we focused on topological terms in sigma-models and did not consider the ones made out of background and dynamic gauge fields. The latter are related to the physics of quantum Hall effect (see \cite{gromov2015framing} for some recent developments) and to the physics of topological insulators and superconductors (e.g., see \cite{ryu2012electromagnetic}). Other interesting but much less understood topics include  topological terms for singular processes (defects in space-time configurations, such as  monopoles, hedgehogs and vortices), the role of topology for phases with gapless fermions, topology and physics out of equilibrium etc. My hope is that these lectures were stimulating enough to encourage the reader to study and to work on applications of topology to physics.

\section{Acknowledgements}

I was inspired to work on topological terms in condensed matter physics by Paul Wiegmann. I have learned most of what is written in these lectures working with Paul and I am infinitely grateful to him for this experience. I would also like to thank Patrick Lee who encouraged my first attempt to lecture on topological terms at MIT. Some of these lectures were shaped in the ``Topics'' course on topological terms for Stony Brook University's physics graduate students. I thank all students and faculty attending those lectures for their attention and questions and to Paul Wiegmann and Artem Abanov for reading the manuscript and providing a very useful feedback. Finally, these notes would never be finished without Somen Bhattacharjee who invited me to SERC School on Topology and Condensed Matter Physics (2015, Kolkata, India) to give these lectures and gently but constantly reminded me to work on the notes.

\bibliographystyle{unsrt}
\bibliography{top-terms}

%

\newpage
\appendix


\section{Appendix: Topological defects and textures}
\la{app:topdefects}

In this appendix we collect few exercises on the topic of homotopy classifications of topological defects and textures in media with spontaneously broken symmetry. A detailed exposure of the topic could be found in the classic reference \cite{1979-Mermin}. Some of known homotopy groups for various spaces are collected in Appendix~\ref{app:homgroups}.

%

{\small

\prn{ex:nematic}
\subsubsection*{Exercise \ref{ex:nematic}: Nematic}

Nematic is a liquid crystal characterized by an order parameter which 
is the unit three-component vector $\mathbf{n}=(n_{1},n_{2},n_{3})$, 
$\mathbf{n}^{2}=1$ with an additional condition $\mathbf{n}\sim -\mathbf{n}$. 
The latter means that two unit vectors which are opposite to each 
other describe the same state. Such an order parameter is called ``director''.

What are the types of topological defects and textures 
allowed for three-dimensional 
nematic? What about two-dimensional one?


\prn{ex:crystal}
\subsubsection*{Exercise \ref{ex:crystal}: Crystal}

One can view a crystalline state as a continuous translational symmetry broken to the subgroup of discrete translations. Then the order parameter space should be identified (for three-dimensional crystal) with $M=G/H = R^{3}/(Z\times Z\times Z)$.
\vspace{0.2cm}

\noindent a) What (geometrically) is the order parameter space for this 
system?
\vspace{0.2cm}

\noindent b) What are the homotopy groups of this manifold 
$\pi_{0,1,2,3}(M)$ ?
\vspace{0.2cm}

\noindent c) What types of topological defects and textures are 
allowed in such a system?
	

\prn{ex:he-3A}
\subsubsection*{Exercise \ref{ex:he-3A}: Superfluid $\mbox{}^{3}He-A$}

The order parameter of superfluid $\mbox{}^{3}He-A$ can be 
represented by two mutually orthogonal unit vectors 
$\mathbf{\Delta}_{1}$, $\mathbf{\Delta}_{2}$. That is, at each point in 
three-dimensional space one has a pair of vectors with properties 
$\mathbf{\Delta}_{1}^{2}=\mathbf{\Delta}_{2}^{2}=1$ and 
$\mathbf{\Delta}_{1}\cdot\mathbf{\Delta}_{2}=0$.
\vspace{0.2cm}

\noindent a) What is the order parameter space for this 
system?
\vspace{0.2cm}

\noindent b) What are the homotopy groups of this manifold 
$\pi_{0,1,2,3}(M)$ ?
\vspace{0.2cm}

\noindent c) What are the types of topological defects and textures 
allowed in such a system?

\prn{ex:heisenberg}
\subsubsection*{Exercise \ref{ex:heisenberg}: Heisenberg  model}

What topological defects and textures should one expect in the ordered state of a three-dimensional classical Heisenberg model? What changes if the order parameter is a director instead of a vector? A ``director'' means a vector without an arrow, i.e., one should identify $\mathbf{S}\equiv -\mathbf{S}$. The models with a director as an order parameter are used to describe nematic liquid crystals (see Ex.~\ref{ex:nematic}).

\prn{ex:XY}
\subsubsection*{Exercise \ref{ex:XY}: Continuum limit of $XY$ model}

Let us start with the $XY$ model defined on a cubic $d$-dimensional lattice. The allowed configurations are parameterized by a planar unit vector $\mathbf{n}_{i}=(\cos\theta_{i},\sin\theta_{i})$ on each site $i$ of the lattice. The energy is given by
\be
	E = -\sum_{\langle ij\rangle}J \cos(\theta_{i}-\theta_{j})\,.
 \la{eq:XYlattice} 
\ee
We assume that the most important configurations are smooth on a lattice scale and one can think of $\theta_{i}$ as of smooth function $\theta(\mathbf{x})$ defined in $R^{d}$ - continuous $d$-dimensional space. Show that the energy is given in this continuous limit by
\be
	E = \frac{J}{2}\int \frac{d^{d}x}{a^{d}}\,  a^{2}(\partial_{\mu}\theta)^{2}\,,
\ee
where $a$ is the lattice constant. The combination $\rho_{s}^{(0)}=J a^{2-d}$ is referred to as \textit{bare spin-wave stiffness} (or \textit{bare superfluid density}). 

Compute the energy of the vortex in such a model. Remember that the divergent integrals should be cut off by lattice constant $a$ and by the size of the system $L$ at small and large distances, respectively. 

For the answer see Ref.~\cite{PolyakovBook-1987}.

\prn{ex:XYcorr}
\subsubsection*{Exercise \ref{ex:XYcorr}: Correlation function $\langle (\theta(x)-\theta(0))^{2}\rangle$}

Calculate the correlation function $\langle (\theta(x)-\theta(0))^{2}\rangle$ in the $XY$ model in $d$ dimensions neglecting the topology of $\theta$, i.e., neglecting vortices and thinking about $\theta$ as of real number without periodicity. Divergencies at small distances should be cut off by the lattice constant $a$. 

Hint: consider $\int d^dq\, \frac{e^{i\mathbf{q}\cdot\mathbf{x}}}{q^2}$ with proper cutoffs. For the answer see Ref.~\cite{PolyakovBook-1987}.


\prn{ex:XYncorrLow}
\subsubsection*{Exercise \ref{ex:XYncorrLow}: Correlation function $\langle\mathbf{n}(x)\cdot\mathbf{n}(0)\rangle$. Low temperatures.}

Using the result of the previous exercise calculate the correlation function $\langle\mathbf{n}(x)\cdot\mathbf{n}(0)\rangle$ in the $XY$ model in $d$ dimensions neglecting the topology of $\theta$. Write  $\langle \mathbf{n}(x)\cdot\mathbf{n}(0)\rangle=\langle \cos(\theta(x)-\theta(0))\rangle
=\R \langle e^{i(\theta(x)-\theta(0))}\rangle$ and use the properties of Gaussian integrals. 

Make the conclusion about the existence of a true long range order in $XY$ model in 2d and relate it to the Mermin-Wagner theorem. 

For the answer see Ref.~\cite{PolyakovBook-1987}.


\prn{ex:XYncorrHigh}
\subsubsection*{Exercise \ref{ex:XYncorrHigh}: Correlation function $\langle\mathbf{n}(x)\cdot\mathbf{n}(0)\rangle$. High temperatures.}

Let us consider high temperatures. Assume that $J/T\ll 1$. Using high temperature expansion for XY-model (\ref{eq:XYlattice}) show that correlation function $\langle\mathbf{n}(x)\cdot\mathbf{n}(0)\rangle$ decays exponentially. Find the correlation length at high temperatures. 

For the answer see Ref.~\cite{PolyakovBook-1987}.

\prn{eq:vortexunbinding}
\subsubsection*{Exercise \ref{eq:vortexunbinding}: Vortex unbinding}

Make an estimate of the BKT phase transition temperature in 2d $XY$ model. Use the energy of the vortex calculated previously, the estimate of the entropy of the vortex, and the condition $F=0$ for the free energy of the vortex.  

For the answer see Ref.~\cite{PolyakovBook-1987}.


}

\section{Appendix: Integrating out $\mathbf{l}$ field}
\la{app:l-derivation}

Consider the $\mathbf{l}$ dependent part of the action (\ref{eq:Sml})
\be
	\int d\tau\,\frac{d^{d}x}{a^d}\,\left( 2J S^{2} a^{2} d\, (\mathbf{l})^{2}
	-Sa^{1}  \mathbf{l} \cdot \Big[\mathbf{h}-\lambda\mathbf{m} 
	+i  d\, (\mathbf{m}\times \p_{\tau}\mathbf{m})\Big]\right) \,.
 \la{eq:action-l}
\ee
Here we added the Lagrange multiplier $\lambda$. The variation with respect to $\lambda$ produces the constraint $\mathbf{l}\cdot\mathbf{m}=0$. The field $\mathbf{l}$ enters the action quadratically and can be integrated out just by substituting $\mathbf{l}$ given by the variational principle. After adding the Lagrange multiplier we can vary over $\mathbf{l}$ without any constraints and obtain
$$
	4J S^{2} a^{2-d} d\, \mathbf{l} -Sa^{1-d}\Big[\mathbf{h} -\lambda \mathbf{m}
	+i  d\, (\mathbf{m}\times \p_{\tau}\mathbf{m})\Big] =0\,
$$
or
$$ 
	\mathbf{l} = \frac{1}{4JSad} \Big[\mathbf{h} -\lambda \mathbf{m}
	+i  d\, (\mathbf{m}\times \p_{\tau}\mathbf{m})\Big] \,.
$$
Substituting the latter expression into (\ref{eq:action-l}) we obtain
\be
	\frac{1}{8Ja^{d}d} \int d\tau\,d^{d}x\, \Big[\mathbf{h} -\lambda \mathbf{m}
	+i  d\, (\mathbf{m}\times \p_{\tau}\mathbf{m})\Big]^{2}\,.
 \la{eq:action-l1}
\ee
Variation over $\lambda$ gives
$$
	 0=\mathbf{m}\cdot \Big[\mathbf{h} -\lambda \mathbf{m}
	+i  d\, (\mathbf{m}\times \p_{\tau}\mathbf{m})\Big] =\mathbf{m}\cdot\mathbf{h} -\lambda
$$
which determines $\lambda = \mathbf{h}\cdot\mathbf{m}$ and gives for the (\ref{eq:action-l1})
\bea
	&-& \frac{1}{8Ja^{d}d} \int d\tau\,d^{d}x\, \Big[\mathbf{h} -\mathbf{m}(\mathbf{h}\cdot\mathbf{m})
	+i  d\, (\mathbf{m}\times \p_{\tau}\mathbf{m})\Big]^{2}
 \nonumber \\
	&= & -\frac{1}{8Ja^{d}d} \int d\tau\,d^{d}x\, \Big[\mathbf{h}^{2} -(\mathbf{h}\cdot\mathbf{m})^{2}
	-  d^{2}\, (\mathbf{m}\times \p_{\tau}\mathbf{m})^{2}
	-2id\, \mathbf{h}\cdot (\mathbf{m}\times \p_{\tau}\mathbf{m})\Big]
 \nonumber \\
	&\to & \frac{1}{8Ja^{d}} \int d\tau\,d^{d}x\, \Big[ 
	 d\, (\p_{\tau}\mathbf{m})^{2}
	+ 2i\, \mathbf{h}\cdot (\mathbf{m}\times \p_{\tau}\mathbf{m}) \Big]\,.
\eea
In the last step we dropped terms independent of $\mathbf{m}$ and the term quadratic in $\mathbf{h}$. Putting all results together, we obtain the effective action which is the functional of the $\mathbf{m}$ field only.
\bea
	S[\mathbf{m}] &=& i \frac{S}{2} \delta_{d,1}
	\int d\tau\,dx\, \mathbf{m}\cdot(\p_{\tau}\mathbf{m}\times \p_{x}\mathbf{m})
 \nonumber \\
	&+& J S^{2} a^{2-d}\frac{1}{2} \int d\tau\,d^{d}x\,
	(\p_{\mu}\mathbf{m})^{2}
	+\frac{d}{8Ja^{d}} \int d\tau\,d^{d}x\, 
	  (\p_{\tau}\mathbf{m})^{2}
 \nonumber \\
 	&+& i\frac{1}{4Ja^{d}} \int d\tau\,d^{d}x\, 
	\mathbf{h}\cdot (\mathbf{m}\times \p_{\tau}\mathbf{m}) \,.
 \nonumber
\eea

\section{Appendix: Homotopy groups often used in physics}
\la{app:homgroups}

In this appendix, we collect some of the homotopy groups often used in physics. Many of these groups can be found in \cite{ito1993encyclopedic}.

\subsection*{Generalities}

If $M$ and $N$ are two topological spaces then for their direct 
product we have
$$\pi_{k}(M\times N) =\pi_{k}(M)\times \pi_{k}(N)\,.$$
If $M$ is a simply-connected topological space 
($\pi_{0}(M)=\pi_{1}(M)=0$) and group $H$ acts on $M$ then 
one can form topological space $M/H$ identifying points of $M$ which 
can be related by some element of $H$ ($x\equiv hx$). Then we have the following relation
$$ \pi_{1}(M/H) =\pi_{0}(H)\,.$$
In particular, if $H$ is a discrete group, $\pi_{0}(H)=H$, and
$$ \pi_{1}(M/H) =H\,.$$
For higher homotopy groups, we have the relation
\be 
	\pi_{k}(M/H)=\pi_{k}(M), \hspace{0.3cm}
	\mbox{if $\pi_{k}(H)=\pi_{k-1}(H)=0\,.$}
 \la{eq:higherhom}
\ee

\subsection*{Homotopy groups of spheres}

For a circle
\begin{eqnarray}
    	\pi_{1}(S^{1})&=&Z, 
 \nonumber \\
    	\pi_{k}(S^{1})&=&0, \hspace{0.3cm}\mbox{for $k\geq 2$}.
 \nonumber
\end{eqnarray}
For higher-dimensional spheres, it is true that
\begin{eqnarray}
   	 \pi_{n}(S^{n})&=&Z, 
 \nonumber \\
    	\pi_{k}(S^{n})&=&0, \hspace{0.3cm}\mbox{for $k< n$}.
 \nonumber
\end{eqnarray}

Homotopy groups of spheres $\pi_{n+k}(S^{n})$ do not depend on $n$ 
for $n>k+1$ (homotopy groups stabilize). In the table below we show in bold face 
the cell from which homotopy groups remain stable (constant along the diagonal of the table).

\begin{center}
\begin{tabular}{|c||c|c|c|c|c|c|c|c|c|c|c|c|}
\hline
    \multicolumn{10}{|c|}{\bfseries Homotopy groups of spheres}
    \\
    \hline
     \mbox{} & $\pi_1$  & $\pi_2$ & $\pi_3$ & $\pi_4$ & 
     $\pi_5$ & $\pi_6$ & $\pi_7$ & $\pi_8$ & $\pi_9$ 
 \\ \hline
    \hline
    $S^{1}$  & $Z$    & $0$   & $0$   & 
    $0$   & $0$  & $0$  & $0$ & $0$ & $0$ 
 \\ \hline
    $S^{2}$  & $0$    & {\blue \bm{$Z$}}   & $Z$   & 
    $Z_{2}$   & $Z_{2}$    & $Z_{12}$   & $Z_{2}$ & 
    $Z_{2}$ & $Z_{3}$ 
 \\ \hline
    $S^{3}$  & $0$    & $0$   & $Z$   & 
    {\blue \bm{$Z_{2}$}}   & $Z_{2}$    & $Z_{12}$   & $Z_{2}$ & 
    $Z_{2}$ & $Z_{3}$ 
 \\ \hline
    $S^{4}$  & $0$    & $0$   & $0$   & 
    $Z$   & $Z_{2}$    & {\blue \bm{$Z_{2}$}}   & $Z\times Z_{12}$ & 
    $Z_{2}\times Z_{2}$ & $Z_{2}\times Z_{2}$ 
 \\ \hline
    $S^{5}$  & $0$    & $0$   & $0$   & 
    $0$   & $Z$    & $Z_{2}$   & $Z_{2}$ & 
    {\blue \bm{$Z_{24}$}} & $Z_{2}$ 
 \\ \hline
    $S^{6}$  & $0$    & $0$   & $0$   & 
    $0$   & $0$    & $Z$   & $Z_{2}$ & 
    $Z_{2}$ & $Z_{24}$ 
 \\ \hline
    $S^{7}$  & $0$    & $0$   & $0$   & 
    $0$   & $0$    & $0$   & $Z$ & 
    $Z_{2}$ & $Z_{2}$ 
 \\ \hline
    $S^{8}$  & $0$    & $0$   & $0$   & 
    $0$   & $0$    & $0$   & $0$ & 
    $Z$ & $Z_{2}$ 
 \\ \hline
\end{tabular}
\end{center}

Here and thereon we denote by $Z$ the group isomorphic to the group of 
integer numbers with respect to addition. $Z_{n}$ is a finite 
Abelian cyclic group. It can be thought of as a group of $n$-th roots 
of unity with respect to a multiplication. Alternatively, it is 
isomorphic to a group of numbers $\{0,1,2,\ldots,n-1\}$ with respect 
to addition modulo $n$. Or simply $Z_{n}=Z/nZ$.

\subsection*{Homotopy groups of Lie groups}

\subsubsection*{Unitary groups}

{\bf Bott periodicity theorem} for unitary groups states that for $k>1$, $n\geq 
\frac{k+1}{2}$
$$
\pi_{k}(U(n)) =\pi_{k}(SU(n))=\left\{
 \begin{array}{ll} 
    0, & \mbox{\hspace{0.6cm} if $k$-even}; \\
    Z, & \mbox{\hspace{0.6cm} if $k$-odd}.
 \end{array}
\right.
$$
The fundamental group $\pi_{1}(SU(n))=0$ and $\pi_{1}(U(n))=1$ for 
all $n$.

In the following table, we show in bold face the entries from which Bott periodicity 
theorem ``starts working'' and table entries become the same further down the column.

\begin{center}
\begin{tabular}{|c||c|c|c|c|c|c|c|c|c|c|c|c|}
\hline
    \multicolumn{13}{|c|}{\bfseries Homotopy groups of unitary groups}
    \\
    \hline
     \mbox{} & $\pi_1$  & $\pi_2$ & $\pi_3$ & $\pi_4$ & 
     $\pi_5$ & $\pi_6$ & $\pi_7$ & $\pi_8$ & $\pi_9$ &
     $\pi_{10}$ & $\pi_{11}$ & $\pi_{12}$ 
 \\ \hline
    \hline
    $U(1)$  & $Z$    & $0$   & $0$   & 
    $0$   & $0$  & $0$  & $0$ & $0$ & $0$ &
    $0$ & $0$ & $0$
 \\ \hline
    $SU(2)$  & $0$    & {\blue \bm{$0$}}   & {\blue \bm{$Z$}}   & 
    $Z_{2}$   & $Z_{2}$    & $Z_{12}$   & $Z_{2}$ & 
    $Z_{2}$ & $Z_{3}$ & $Z_{15}$ & $Z_{2}$ & $Z_{2}\times Z_{2}$ 
 \\ \hline
    $SU(3)$  & $0$    & $0$   & $Z$   & 
    {\blue \bm{$0$}}   & {\blue \bm{$Z$}}    & $Z_{6}$   & $0$ & 
    $Z_{12}$ & $Z_{3}$ & $Z_{30}$ & $Z_{4}$ & $Z_{60}$ 
 \\ \hline
    $SU(4)$  & $0$    & $0$   & $Z$   & 
    $0$   & $Z$    & {\blue \bm{$0$}}   & {\blue \bm{$Z$}} & 
    $Z_{24}$ & $Z_{2}$ & $Z_{120}\times Z_{2}$ &  $Z_{4}$ & $Z_{60}$ 
 \\ \hline
    $SU(5)$  & $0$    & $0$   & $Z$   & 
    $0$   & $Z$    & $0$   & $Z$ & 
    {\blue \bm{$0$}} & {\blue \bm{$Z$}} & $ $ &  $ $ & $ $ 
 \\ \hline
\end{tabular}
\end{center}

\subsubsection*{Orthogonal groups}

{\bf Bott periodicity theorem} for orthogonal groups states that for $n\geq 
k+2$
$$
\pi_{k}(O(n)) =\pi_{k}(SO(n))=\left\{
 \begin{array}{ll} 
    0, & \mbox{\hspace{0.6cm} if $k=2,4,5,6\;\;(mod\; 8)$}; \\
    Z_{2}, & \mbox{\hspace{0.6cm} if $k=0,1\;\;(mod\; 8)$}; \\
    Z, & \mbox{\hspace{0.6cm} if $k=3,7\;\;(mod\; 8)$}.
 \end{array}
\right.
$$

In the following table, we we show in bold face the entries from which Bott periodicity 
theorem ``starts working''.

\begin{center}
\begin{tabular}{|l||c|c|c|c|c|c|c|c|c|c|c|c|}
\hline
    \multicolumn{9}{|c|}{\bfseries Homotopy groups of orthogonal 
    groups}
    \\
    \hline
     \mbox{} & $\pi_1$  & $\pi_2$ & $\pi_3$ & $\pi_4$ & 
     $\pi_5$ & $\pi_6$ & $\pi_7$ & $\pi_8$ 
 \\ \hline
    \hline
    $SO(2)$  & $Z$    & $0$   & $0$   & 
    $0$   & $0$  & $0$  & $0$ & $0$ 
 \\ \hline
    $SO(3)$  & {\blue \bm{$Z_{2}$}}    & $0$   & $Z$   & 
    $Z_{2}$   & $Z_{2}$    &  $Z_{12}$  & $Z_{2}$ & 
    $Z_{2}$ 
 \\ \hline
    $SO(4)$  & $Z_{2}$    & {\blue \bm{$0$}}   & $(Z)^{\times 2}$   & 
    $(Z_{2})^{\times 2}$   & $(Z_{2})^{\times 2}$    & $(Z_{12})^{\times 2}$ & $(Z_{2})^{\times 2}$ & 
    $(Z_{2})^{\times 2}$ 
 \\ \hline
    $SO(5)$  & $Z_{2}$    & $0$   & {\blue \bm{$Z$}}   & 
    $Z_{2}$   & $Z_{2}$    & $0$   & $Z$ & 
    $0$ 
 \\ \hline
    $SO(6)$  & $Z_{2}$    & $0$   & $Z$   & 
    {\blue \bm{$0$}}   & $Z$    & $0$   & $Z$ & 
    $Z_{24}$ 
 \\ \hline
    $SO(n>6)$  & $Z_{2}$    & $0$   & $Z$   & 
    $0$   & {\blue \bm{$0$}}    & $0$   & $ $ & 
    $ $ 
 \\ \hline
\end{tabular}
\end{center}

\subsubsection*{Symplectic groups}

{\bf Bott periodicity theorem} for symplectic groups states that for $n\geq 
\frac{k-1}{4}$
$$
\pi_{k}(Sp(n)) =\left\{
 \begin{array}{ll} 
    0, & \mbox{\hspace{0.6cm} if $k=0,1,2,6\;\;(mod\; 8)$}; \\
    Z_{2}, & \mbox{\hspace{0.6cm} if $k=4,5\;\;(mod\; 8)$}; \\
    Z, & \mbox{\hspace{0.6cm} if $k=3,7\;\;(mod\; 8)$}.
 \end{array}
\right.
$$

In the following table, we we show in bold face the entries from which Bott periodicity 
theorem ``starts working''.

\begin{center}
\begin{tabular}{|c||c|c|c|c|c|c|c|c|c|c|c|c|}
\hline
    \multicolumn{13}{|c|}{\bfseries Homotopy groups of symplectic 
    groups}
    \\
    \hline
     \mbox{} & $\pi_1$  & $\pi_2$ & $\pi_3$ & $\pi_4$ & 
     $\pi_5$ & $\pi_6$ & $\pi_7$ & $\pi_8$ & $\pi_9$ &
     $\pi_{10}$ & $\pi_{11}$ & $\pi_{12}$ 
 \\ \hline
    \hline
    $Sp(1)$  & $0$    & {\blue \bm{$0$}}   & {\blue \bm{$Z$}}   & 
    {\blue \bm{$Z_{2}$}}   & {\blue \bm{$Z_{2}$}}    & $Z_{12}$   & $Z_{2}$ & 
    $Z_{2}$ & $Z_{3}$ & $Z_{15}$ & $Z_{2}$ & $Z_{2}\times Z_{2}$ 
 \\ \hline
    $Sp(2)$  & $0$    & $0$   & $Z$   & 
    $Z_{2}$   & $Z_{2}$    &  {\blue \bm{$0$}}  & {\blue \bm{$Z$}} & 
    {\blue \bm{$0$}} & {\blue \bm{$0$}} & $Z_{120}$ & $Z_{2}$ & 
    $Z_{2}\times Z_{2} $ 
 \\ \hline
    $Sp(n\geq 3)$  & $0$    & $0$   & $Z$   & 
    $Z_{2}$   & $Z_{2}$    &  $0$  & $Z$ & 
    $0$ & $0$ & {\blue \bm{$0$}} & {\blue \bm{$Z$}} & {\blue \bm{$Z_{2}$}} 
 \\ \hline
\end{tabular}
\end{center}

\subsubsection*{Exceptional groups}

\begin{center}
\begin{tabular}{|l||c|c|c|c|c|c|c|c|c|c|c|c|}
\hline
    \multicolumn{13}{|c|}{\bfseries Homotopy groups of exceptional 
    groups}
    \\
    \hline
     \mbox{} & $\pi_1$  & $\pi_2$ & $\pi_3$ & $\pi_4$ & 
     $\pi_5$ & $\pi_6$ & $\pi_7$ & $\pi_8$ & $\pi_9$ & $\pi_{10}$ & 
     $\pi_{11}$ & $\pi_{12}$
 \\ \hline
    \hline
    $G_{2}$  & $0$    & $0$   & $Z$   & 
    $0$   & $0$    & $Z_{3}$  & $0$ & $Z_{2}$ & $Z_{6}$ & $0$ & 
    $Z\times Z_{2}$ & $0$ 
 \\ \hline
    $F_{4}$  & $0$    & $0$   & $Z$   & 
    $0$   & $0$    &  $0$  & $0$ & $Z_{2}$ & $Z_{2}$ & $0$ & 
    $Z\times Z_{2}$ & $0$
 \\ \hline
    $E_{6}$  & $0$    & $0$   & $Z$   & 
    $0$   & $0$    &  $0$  & $0$ & $0$ & $Z$ & $0$ & 
    $Z$ & $Z_{12}$ 
 \\ \hline
    $E_{7}$  & $0$    & $0$   & $Z$   & 
    $0$   & $0$    &  $0$  & $0$ & $0$ & $0$ & $0$ & 
    $Z$ & $Z_{2}$
 \\ \hline
    $E_{8}$  & $0$    & $0$   & $Z$   & 
    $0$   & $0$    &  $0$ & $0$ & $0$ & $0$ & $0$ & 
    $0$ & $0$ 
 \\ \hline
\end{tabular}
\end{center}

\subsection*{Homotopy groups of some other spaces}

\subsubsection*{Tori}

An $n$-dimensional torus can be defined as a direct product of $n$ 
circles $T^{n}=(S^{1})^{\times n}$. One can immediately derive that 
\begin{eqnarray}
    	\pi_{1}(T^{n})&=&(Z)^{\times n}\,,
 \nonumber \\
    	\pi_{k}(T^{n})&=&0, \hspace{0.3cm}\mbox{for $k\geq 2$}\,.
 \nonumber
\end{eqnarray}
    
\subsubsection*{Projective spaces}

The real projective space $RP^{n}$ can be represented as 
$RP^{n}=S^{n}/Z_{2}$. For $n=1$  $RP^{1}=S^{1}$. We have from (\ref{eq:higherhom})
\begin{eqnarray}
   	\pi_{1}(RP^{1})&=&Z\,, 
 \nonumber \\
   	\pi_{1}(RP^{n})&=&Z_{2}, \hspace{0.3cm}\mbox{for $n\geq 2$}\,,
 \nonumber \\
   	\pi_{k}(RP^{n})&=&\pi_{k}(S^{n}),
   	\hspace{0.3cm}\mbox{for $k\geq 2$}\,. 
 \nonumber   
\end{eqnarray}

\begin{center}
\begin{tabular}{|c||c|c|c|c|c|c|c|c|c|c|c|c|}
\hline
    \multicolumn{10}{|c|}{\bfseries Homotopy groups of real 
    projective spaces }
    \\
    \hline
     \mbox{} & $\pi_1$  & $\pi_2$ & $\pi_3$ & $\pi_4$ & 
     $\pi_5$ & $\pi_6$ & $\pi_7$ & $\pi_8$ & $\pi_9$ 
 \\ \hline
    \hline
    $RP^{1}$  & $Z$    & $0$   & $0$   & 
    $0$   & $0$    & $0$   & $0$ & 
    $0$ & $0$ 
 \\ \hline
    $RP^{2}$  & $Z_{2}$    & $Z$   & $Z$   & 
    $Z_{2}$   & $Z_{2}$    & $Z_{12}$   & $Z_{2}$ & 
    $Z_{2}$ & $Z_{3}$ 
 \\ \hline
    $RP^{3}$  & $Z_{2}$    & $0$   & $Z$   & 
    $Z_{2}$   & $Z_{2}$    & $Z_{12}$   & $Z_{2}$ & 
    $Z_{2}$ & $Z_{3}$ 
 \\ \hline
    $RP^{4}$  & $Z_{2}$    & $0$   & $0$   & 
    $Z$   & $Z_{2}$    & $Z_{2}$   & $Z\times Z_{12}$ & 
    $Z_{2}\times Z_{2}$ & $Z_{2}\times Z_{2}$ 
 \\ \hline
\end{tabular}
\end{center}

Similarly, for complex projective spaces $CP^{n}$ we have 
$CP^{1}=S^{2}$ and, generally, $CP^{n}=S^{2n+1}/S^{1}$. We have for 
homotopy groups
\begin{eqnarray}
   \pi_{1}(CP^{n})&=&0, 
 \nonumber \\
   \pi_{2}(CP^{n})&=&Z, 
 \nonumber \\
   \pi_{k}(CP^{n})&=&\pi_{k}(S^{2n+1}),
   \hspace{0.3cm}\mbox{for $k\geq 3$}. 
 \nonumber   
\end{eqnarray}

\begin{center}
\begin{tabular}{|c||c|c|c|c|c|c|c|c|c|c|c|c|}
\hline
    \multicolumn{13}{|c|}{\bfseries Homotopy groups of complex 
    projective spaces }
    \\
    \hline
     \mbox{} & $\pi_1$  & $\pi_2$ & $\pi_3$ & $\pi_4$ & 
     $\pi_5$ & $\pi_6$ & $\pi_7$ & $\pi_8$ & $\pi_9$ &
     $\pi_{10}$ & $\pi_{11}$ & $\pi_{12}$ 
 \\ \hline
    \hline
    $CP^{1}$  & $0$    & $Z$   & $Z$   & 
    $Z_{2}$   & $Z_{2}$    & $Z_{12}$   & $Z_{2}$ & 
    $Z_{2}$ & $Z_{3}$ & $Z_{15}$ & $Z_{2}$ & $Z_{2}\times Z_{2}$ 
 \\ \hline
    $CP^{2}$  & $0$    & $Z$   & $0$   & 
    $0$   & $Z$    & $Z_{2}$   & $Z_{2}$ & 
    $Z_{24}$ & $Z_{2}$ & $Z_{2}$ & $Z_{2}$ & $Z_{30}$ 
 \\ \hline
    $CP^{3}$  & $0$    & $Z$   & $0$   & 
    $0$   & $0$    & $0$   & $Z$ & 
    $Z_{2}$ & $Z_{2}$ & $Z_{24}$ & $0$ & $0$
 \\ \hline
    $CP^{4}$  & $0$    & $Z$   & $0$   & 
    $0$   & $0$    & $0$   & $0$ & 
    $0$ & $Z$ & $Z_{2}$ & $Z_{2}$ & $Z_{24}$
 \\ \hline
\end{tabular}
\end{center}

\end{document}